\documentclass[twocolumn,amsmath,amssymb,superscriptaddress,aps,pra]{revtex4-2}

\usepackage{color}
\usepackage{graphicx}
\usepackage{amssymb}
\usepackage{amsmath}

\newcommand{\mynomencl}[3][subsection]{%
  \begingroup\edef\x{\endgroup
  \unexpanded{\nomenclature{#2}}%
    {\unexpanded{#3} (\csname the#1\endcsname)}}\x} 

\newcommand{\ud}{\mathrm{d}}

\let\Eps\varepsilon

\usepackage[breaklinks=true]{hyperref}
\graphicspath{ {./figures/} {./}}

\begin{document}

\title{Stochastic modulational instability in the nonlinear Schr\"odinger equation with colored random dispersion}

\author{Andrea Armaroli} 

\affiliation{Univ.~Lille,  CNRS,  UMR  8523-PhLAM-Physique  des  Lasers  Atomes  et  Mol\'ecules,  F-59000  Lille,  France}

\author{Guillaume Dujardin} 

\affiliation{Univ.~Lille, Inria, CNRS, UMR 8524 - Laboratoire Paul Painlev{\'e}, F-59000 Lille, France}

\author{Alexandre Kudlinski} 
\affiliation{Univ.~Lille,  CNRS,  UMR  8523-PhLAM-Physique  des  Lasers  Atomes  et  Mol\'ecules,  F-59000  Lille,  France}

\author{Arnaud Mussot} 

\affiliation{Univ.~Lille,  CNRS,  UMR  8523-PhLAM-Physique  des  Lasers  Atomes  et  Mol\'ecules,  F-59000  Lille,  France}

\author{Stefano Trillo} 

\affiliation{Department  of  Engineering,  University  of  Ferrara,  I-44122  Ferrara,  Italy}

\author{Stephan De Bi{\`e}vre} 

\affiliation{Univ.~Lille, CNRS, Inria, UMR 8524 - Laboratoire Paul Painlev{\'e}, F-59000 Lille, France}

\author{Matteo Conforti} 

\affiliation{Univ.~Lille,  CNRS,  UMR  8523-PhLAM-Physique  des  Lasers  Atomes  et  Mol\'ecules,  F-59000  Lille,  France}

\begin{abstract}

We study modulational instability (MI) in optical fibers with random group-velocity dispersion (GVD). We consider Gaussian and dichotomous \textit{colored} stochastic processes.  We resort to different analytical methods (namely, the cumulant expansion and the functional approach) and assess their reliability in estimating the MI gain of stochastic origin. If the power spectral density (PSD) of the GVD fluctuations is centered at null wavenumber, we obtain low-frequency MI sidelobes which converge to those given by a white noise perturbation when the correlation length tends to $0$. If instead the stochastic processes are  modulated in space, one or more MI sidelobe pairs corresponding to the well-known parametric resonance (PR) condition can be found. A transition from small and broad sidelobes to peaks nearly indistinguishable from PR-MI is predicted, in the limit of large perturbation amplitudes and correlation lengths of the random process. We find that the cumulant expansion provides good analytical estimates for small PSD values and small correlation lengths, when the MI gain is very small. The functional approach is rigorous only for the dichotomous processes, but allows us to model a wider range of parameters and to predict the existence of MI sidelobes comparable to those observed in homogeneous fibers of anomalous GVD.

\end{abstract}

\maketitle

\section{Introduction}

A physical system exhibiting an interplay of weak nonlinearity and group velocity dispersion (GVD) is subject to modulational instability (MI), \textit{i.e.}, the destabilization of a homogeneous state (plane or continuous waves), via the exponential growth of small harmonic perturbations on a uniform background \cite{Zakharov2009}. After pioneering works in fluid mechanics \cite{Benjamin1967,Zakharov1968}, MI was discovered in electromagnetic waves \cite{Bespalov1966} as well as in plasmas \cite{H.Ichikawa1973}; in the 80s the phenomenon was observed in nonlinear fiber optics \cite{Tai1986}.
In uniform fibers, MI arises for anomalous (negative) GVD, but it may also appear for normal GVD if polarization \cite{Berkhoer1970}, higher-order modes \cite{Stolen1974} or higher-order dispersion are considered \cite{Cavalcanti1991}.
A different kind of MI related to a parametric resonance (PR) mechanism emerges when the dispersion or the nonlinearity of the fiber are periodically modulated
\cite{Smith1996,Droques2012,Armaroli2012,Mussot2018}. 

The impact of a random variation of GVD on MI was also the subject of a considerable research effort.
The particular case in which the GVD is perturbed by a Gaussian white noise, i.e., a process exhibiting a vanishing correlation length or equivalently a flat power spectral density (PSD), is exactly solvable\cite{Abdullaev1996,Abdullaev1997,Abdullaev1999,Garnier2000,Chertkov2001}. When the unperturbed fiber has an anomalous GVD, the conventional MI gain profile is deformed due to the random perturbation. In addition,  MI sidebands of stochastic origin appear in the case of normal GVD. A white noise, however, implies arbitrarily large variations of GVD  on arbitrarily small scales: an idealization that does not always provide a relevant modeling of the randomness that may occur in physical fibers.
A non-conclusive theoretical study of parametric amplification in the case of a GVD perturbed by a Gaussian process with a finite correlation length was proposed in \cite{Karlsson1998},
and a numerical study can be found in \cite{Farahmand2004}.

We aim at studying the MI problem  in a class of random-GVD fibers that is both experimentally accessible and theoretically tractable. 
In \cite{Dujardin2021}, we studied the case of a GVD perturbed by  randomly located sharp and large \textit{kicks}. Two different families of random  processes were chosen to generate their mutual spacing and amplitude. Different MI sidebands were predicted, including multibump ones around zero detuning and others localized around PR frequencies. 

Here, we consider random fluctuations extended in space as in Ref.~\cite{Abdullaev1996}, but focus on \textit{colored} processes exhibiting an exponentially decaying autocorrelation function. A lowpass (LP) and a bandpass (BP, modulated) case are considered. 
We resort to cumulant expansion \cite{Bourret1962,vanKampen1976,vanKampen1974a,vanKampen1974, Terwiel1974,Fox1986,vanKampenBook}, functional methods \cite{Furutsu1963,Novikov1965,Shapiro1978}, and numerical simulations. The LP processes converge to the white noise results for vanishing correlation lengths. The MI sidelobes are located in the same detuning range and have amplitudes of the same order of magnitude of those calculated from white noise. The BP processes yield PR-like MI sidebands that are also comparable (in their maximal values) to the white noise MI gain. 
 For this reason, the white noise turns out to be a reference for many crucial
  properties of the MI gain in a more generic setting, eventhough there is little
  hope of an experimental realization of it.


Since the values of frequency detuning span a large range, the validity of the different analytical methods has to be questioned, like in other physical settings and systems \cite{Zhang1992,GittermanNoisy}. We thus have to rely on different approaches and comparatively assess their soundness in describing the different features of the MI sidelobes (position, height, width). 

The rest of the paper is organized as follows. After presenting the model equations (Sec. \ref{sec:model}), we study in Sec.~\ref{sec:LP} the lowpass random fluctuation by means of the two different approaches mentioned above. We then analyze in Sec.~\ref{sec:BP} bandpass random fluctuations, for which we find convenient to introduce an averaging approach to the stochastic equations. In each case we present a thorough comparison of analytical approaches to numerical results. To close the paper we report our conclusions and perspectives.

\section{Model description}
\label{sec:model}

The propagation of optical pulses in a nonlinear optical fiber can be modeled by the universal nonlinear Schr{\"o}dinger equation (NLSE),
\begin{equation}
   i\partial_z U - \frac{1}{2}\beta_2(z)\partial_{tt}U  + \gamma |U|^2U = 0.
\label{eq:NLS1}
\end{equation}
Here $U(z,t)$ is the envelope of the optical pulse field in units of $\sqrt{W}$, function of  the propagation distance $z$ and time $t$ in the frame of reference propagating at the group velocity of the mode propagating in the fiber, $\beta_2$ is the GVD and $\gamma$ is the nonlinear coefficient \cite{Agrawal2012}. We suppose that $\beta_2$ fluctuates randomly in $z$, while $\gamma$ stays constant.

Let $\beta_2(z) = \beta_2^0 +\delta\!\beta(z)$, where $\beta_2^0>0$ (normal average GVD) and $\delta\!\beta(z)$ is a stochastic process of zero mean, that we specify below. 

We observe that, for $\delta\!\beta=0$, Eq.~\eqref{eq:NLS1} has a continuous wave ($t$-independent) solution $U_0(z) = \sqrt{P}\exp\left(i \gamma P z\right)$.

In order to study the stability of this continuous wave solution, we insert in  Eq.~\eqref{eq:NLS1} the perturbed solution   ${U}(z,t) = \left[\sqrt{P} + \check x_1(z,t) + i \check x_2(z,t)\right]\exp (i \gamma P z)$, linearize and Fourier-transform the resulting equation with respect to $t$ ($\omega$ is used as the associated angular frequency detuning from the carrier $U_0$), to obtain  
\begin{equation}
   \frac{\ud x}{\ud z}
	 	 = 
	 	\begin{bmatrix}
	 		0 & -g(z) \\h(z) & 0
	 	\end{bmatrix}
	 	x,
\label{eq:MIeqs1}
\end{equation}
with $x  \equiv ( x_1,x_2)^\mathrm{T}$ (functions of $ \omega$ and $z$), $g(z) = g_0+\delta\!g(z)  $ and $h(z) = h_0 + \delta\!g(z) $, with $g_0 \equiv \beta_2^0 \frac{\omega^2}{2} $, $h_0\equiv  g_0 + 2\gamma P$, and $\delta\! g(z)\equiv {\delta\!\beta (z)}\frac{\omega^2}{2}$. 
Eq.~\eqref{eq:MIeqs1} is a system of stochastic differential equations (SDEs) for each value $\omega$. In the following sections we will discuss how to reduce it to a system of ordinary differential equations (ODEs) for the first and second moments of the probability density function of $x$ in order to estimate the MI gain.

We recall that we consider only normal average GVD, thus 
\begin{equation}
k^2\equiv g_0h_0>0
\end{equation}
and no conventional MI appears for $\delta\!\beta (z)=0$. The MI sidebands we predict below are therefore completely ascribed to random fluctuations. The case of anomalous average GVD will be the subject of future work.


We focus here on two families of random processes, where the stochastic process $\delta\!\beta$ is characterized by two parameters $N_0>0$ and $B>0$.
The first family, for which we denote $\delta\!\beta\equiv\chi$, is characterized by an autocorrelation function of the form
\begin{equation}
 R_\chi(\zeta\equiv z-z') \equiv \langle \chi(z)\chi(z')\rangle = \frac{N_0B}{4}\exp(-B|\zeta|).
\label{eq:Rchi}
 \end{equation}
We recall that the variance of the process is $\sigma_\chi^2 = R_\chi(0) = \frac{N_0B}{4}$. For $B\to \infty$, $R_\chi\to \frac{N_0}{2}\delta(\zeta)$, {\it i.e.}, the white noise autocorrelation function.
Different stochastic processes exhibit this same autocorrelation function. Here, we consider the Gaussian (often denoted in the physics literature as Ornstein-Uhlenbeck \cite{vanKampenBook}) and the dichotomous processes, which find important applications in physics and allow us to obtain workable approximations \cite{GittermanNoisy}.

Both are stationary in $z$. 
By virtue of the Wiener-Khinchin theorem, the PSD of $\chi$ coincides with the Fourier transform of $R_\chi$
\begin{equation}
S_\chi(\kappa) = \int\limits_{-\infty}^\infty {\ud\,\zeta} R_\chi(\zeta) \exp(i\kappa \zeta) = \frac{N_0}{2} \frac{B^2}{B^2 + \kappa^2}.
\label{eq;PSDdef}
\end{equation}

The Gaussian process is numerically generated in the  $\kappa$-domain  by filtering an approximated white noise of PSD $N_0/2$ by means of a \textit{lowpass} filter of transfer function 
\begin{equation}
H_\mathrm{LP}(\kappa) = \frac{B}{B+i\kappa}.
\label{eq:LorLP}
\end{equation}
The dichotomous process is more conveniently obtained directly in the $z$ domain by switching the amplitude of the fluctuation between $\pm \sigma_\chi$ with an exponentially distributed \emph{waiting} (i.e., between switching points) length with mean $2/B$. Both are pertinent to fiber optics.
Indeed, the Gaussian process corresponds to a continuous variation of dispersion as can be obtained by varying the fiber radius during the drawing process. The dichotomous process corresponds to splicing together fibers with  different GVD $\beta_2 = \beta_2^0 (1\pm \sigma _\chi)$ and random lengths. We refer to the processes belonging to this class, with autocorrelation function as in Eq.~\eqref{eq:Rchi} and PSD as in Eq.~\eqref{eq:LorLP} as LP processes.

The second family of stochastic processes, for which we denote $\delta\!\beta\equiv\xi$, is obtained as the modulated version of $\chi$ with central wavenumber $\kappa_0=\frac{2\pi}{ \Lambda_0}>0$ ($\Lambda_0$ is the associated spatial period), that is written in phase-quadrature representation
\begin{equation}
\xi(z) = \psi_1(z) \cos \kappa_0 z + \psi_2(z) \sin  \kappa_0 z.
\label{eq:PQxi}
\end{equation}
Here $\psi_{1,2}$ two stationary (in $z$) and independent random processes with zero mean and autocorrelation functions
\begin{equation}
 R_{\psi_i}(\zeta) =   \frac{N_0B}{2} \exp(-B|\zeta|),
 \label{eq:Rpsi}
\end{equation}
for $i=1,2$; moreover, $\langle\psi_i\psi_j\rangle = \frac{N_0B}{2}\delta_{ij}$. 
In analogy to $\chi$, we consider
either two Gaussian or two dichotomous processes for $\psi_{1,2}$. They are generated according to their distribution as is done for $\chi$.
The process $\xi$ thus exhibits an autocorrelation function of the form
\begin{equation}
 R_\xi(\zeta) = \frac{N_0B}{2}\cos \kappa_0\zeta \exp(-B|\zeta|).
\label{eq:Rxi}
 \end{equation}
The variance of the process is $\sigma^2_\xi = \frac{N_0B}{2}$, like for $\psi_{1,2}$.

The PSD of $\xi$ reads
\begin{equation}
  S_\xi(\kappa) = \frac{N_0}{2} \left[\frac{B^2}{B^2 + (\kappa-\kappa_0)^2} + \frac{B^2}{B^2 + (\kappa+\kappa_0)^2}\right].
  \label{eq:LorBP}
\end{equation}
We note that, for $B\ll \kappa_0$, $S_\xi$ is centered approximately around the wavenumber $\pm \kappa_0$, with $S_\xi(\pm \kappa_0)\approx \frac{N_0}{2}=S_\chi(0)$  and  bandwidth $B$ (in wavenumber units). We refer to this family  with autocorrelation function as in Eq.~\eqref{eq:Rxi} and PSD as in Eq.~\eqref{eq:LorBP} as BP processes.

For both LP and BP families,  we  employ the definition of correlation length \cite{StranovichBook163}
$
\zeta^\mathrm{c} \equiv \frac{1}{R(0)}\int\limits_0^\infty{\ud \zeta |R(\zeta)}| 
$,
 which gives  $\zeta_\chi^\mathrm{c} = 1/B$ for $\chi$ and $\zeta^\mathrm{c}_\xi\approx 2/(\pi B)$ for $\xi$, if $B \ll \kappa_0$. 

In the next two sections we will study the effect $\chi$ and $\xi$, respectively, on the MI predicted by Eq.~\eqref{eq:MIeqs1}. 

\section{Lowpass random dispersion}
\label{sec:LP}

First we consider processes with the autocorrelation function given in Eq.~\eqref{eq:Rchi}. We will discuss both the first and the second moment equations associated to Eq.~\eqref{eq:MIeqs1}.

\subsection{Cumulant expansion (first moments)}

The cumulant expansion yields a series development for the ODEs associated to a SDE \cite{vanKampenBook}. It is similar to the Dyson series of scattering theory \cite{Dubkov1977} and provides a solid base for more \textit{ad-hoc} schemes \cite{Bourret1962}. See \cite{Terwiel1974} for a systematic derivation of terms to arbitrary order.

Let us rewrite Eq.~\eqref{eq:MIeqs1} in the standard form 
$\dot{ {x}} = \left[ {A_1} + \alpha\eta(z)  {C_1} \right] {x}$ with
\[
 {A_1} = \begin{bmatrix}
	 		0 & -g_0 \\h_0 & 0
	 	\end{bmatrix}, \;
 {C_1} = \begin{bmatrix}
	 		0 & -1 \\1 & 0
	 	\end{bmatrix}, 	 	
\]
 $\alpha = \frac{\omega^2}{4}\sqrt{N_0 B}$ and $\eta = 2\chi/\sqrt{N_0 B}$ a random process with unit variance and zero mean.
The expansion is performed in the formal parameter $\alpha$. To second order (the first-order term is obviously 0), we write the ODE for the first moment $\langle  {x}\rangle = (\langle 
x_1\rangle, \langle 
x_2\rangle)^\mathrm{T}$  as 
\begin{equation}
 \frac{\ud}{\ud z}\langle {x}\rangle = 
	 \left[\mathrm{A_1} + \alpha^2 K_2^1\right]\langle {x}\rangle,
	 \label{eq:CumExp}
\end{equation}
with 
\begin{equation}
K_2^1 = \int\limits_{0}^{\infty}\!\ud \zeta \, C_1 e^{A_1\zeta} C_1e^{-A_1\zeta} R_\eta(\zeta).
\label{eq:K21}
\end{equation}

Other terms can be added in the expansion of Eq.~\eqref{eq:CumExp}: their contribution to the solution rapidly decreases if
\begin{equation}
  \label{eq:defvarepsilon}
  \varepsilon\equiv\alpha\zeta^\mathrm{c},
\end{equation}
is small, {\it i.e.} $\frac{\omega^2}{4}\sqrt{\frac{N_0}{B}}\ll 1$. For large detuning or small filter bandwidth $B$, the approximation may be invalid. A fixed initial condition $x(0)$ can be considered and the solutions of Eq.~\eqref{eq:CumExp} do not keep memory of it for $z\gg\zeta_\mathrm{c}$. The long term dynamics being our main focus, we set the limit of integration  to infinity in Eq.~\eqref{eq:K21}, see \cite{vanKampen1974}.
 
By tedious but straightforward algebra, we obtain
\begin{widetext}
\begin{equation}
  \frac{\ud}{\ud z}
	 	\langle {x}\rangle =
	 	\begin{bmatrix}
	 		-\frac{\omega^4}{4}\frac{g_0}{2k^2}\left[(g_0+h_0)c_1 -  (h_0-g_0)c_2\right] &  -g_0 +\frac{\omega^4}{8k} (h_0-g_0)c_3\\
 	 		h_0 + \frac{\omega^4}{8k} (h_0-g_0) c_3 & 
 	 		-\frac{\omega^4}{4}\frac{h_0}{2k^2}\left[(g_0+h_0)c_1 + (h_0-g_0) c_2\right]
	 	\end{bmatrix}
	 	\langle {x}\rangle,
\label{eq:M1direct2}
\end{equation}
\end{widetext}
with
\begin{equation}
\begin{aligned}
c_1&\equiv \int \limits_0^{\infty}\ud \zeta R_\chi(\zeta) = \frac{1}{2}S_\chi(0)=\frac{N_0}{4},\\ 
c_2 & \equiv \int \limits_0^{\infty}\ud \zeta R_\chi(\zeta) \cos 2k \zeta= \frac{1}{2}S_\chi(2k)= \frac{N_0B^2}{4} \frac{1}{B^2 + 4k^2}
\\
c_3 &\equiv \int\limits_0^\infty \ud\zeta R_\chi(\zeta)\sin(2k\zeta) = \frac{N_0 Bk}{2} \frac{1}{B^2 + 4k^2}.
\end{aligned}
\end{equation} 
Since $c_i>0$, $i=1,2,3$, for all $\omega$, and $c_1>c_2$, it is easy to verify that the eigenvalues of the matrix in Eq.~\eqref{eq:M1direct2} have both a negative real part, so that the system Eq.~\eqref{eq:M1direct2} does not predict any MI gain. This finding is analogous to the conventional harmonic oscillator with random frequency \cite{vanKampenBook, GittermanNoisy}, for which the first moment undergoes only damping. For this reason it is necessary to resort to the equations for the second moments.

We recall that, for white noise, the cumulant expansion at second order is exact and in that limit, $B\to\infty$, $c_1=c_2$ and $c_3 = 0$, so the eigenvalues of Eq.~\eqref{eq:M1direct2} reduce to Eq.~(23) of Ref.~\cite{Abdullaev1996}.

\subsection{Cumulant expansion (second moments)}
\label{ssec:cumexp2}

We now consider second moments. 
First we let $X_1 = x_1^2$, $X_2 = x_2^2$, and $X_3 = x_1x_2$ and derive from Eq.~\eqref{eq:MIeqs1} a system for their  evolution, which reads
\begin{equation}
	 	 	\frac{\ud}{\ud z}
	 	 {X} = 
	 	\begin{bmatrix}
	 		0 & 0& -2 g(z) \\
	 		0 & 0& 2 h(z) \\
	 		h(z) & -g(z) & 0
	 	\end{bmatrix}
	 	 {X}.
	 	\label{eq:MIeqs2}
\end{equation}

In order to perform the cumulant expansion, we write Eq.~\eqref{eq:MIeqs2} in standard form by letting
\[
 {A_2} = \begin{bmatrix}
	 		0 & 0 & -2g_0 \\0 & 0 & 2h_0 \\ h_0 & -g_0 & 0 
	 	\end{bmatrix}, \;
 {C_2} = \begin{bmatrix}
	 		0 & 0&  -2 \\ 0 & 0&  2 \\1 & -1 & 0 
	 	\end{bmatrix}, 	 	
\]
the other quantities  $\alpha$ and $\eta$ being the same as in the previous subsection. 

Up to second order, the cumulant expansion reads
\begin{equation}
\begin{gathered}
 \frac{\ud}{\ud z}\langle {X}\rangle = 
	 \left[\mathrm{A_2} + \alpha^2 K_2^2\right]\langle {X}\rangle, \\\text{with }
K_2^2 = \int\limits_{0}^{\infty}\!\ud \zeta \, C_2 e^{A_2\zeta} C_2e^{-A_2\zeta} R_\eta(\zeta),
\end{gathered}
	 \label{eq:CumExp2}
\end{equation}
which gives
\begin{widetext}
\begin{equation}
  \frac{\ud}{\ud z}
	 	\langle {X}\rangle  
	 	= 
	 	\begin{bmatrix}
		-\frac{\omega^4}{4g_0}\left[(g_0+h_0) c_1 + (g_0-h_0)c_2\right] & \frac{\omega^4}{4h_0}\left[(g_0+h_0)c_1 - (g_0-h_0)c_2\right] & -2g_0 \\
		\frac{\omega^4}{4g_0}\left[(g_0+h_0)c_1 + (g_0-h_0)c_2\right] & -\frac{\omega^4}{4h_0}\left[(g_0+h_0)c_1 - (g_0-h_0)c_2\right] &2 h_0 \\
		h_0+\frac{\omega^4}{4k}(h_0-g_0)c_3 & -g_0+\frac{\omega^4}{4k}(h_0-g_0)c_3 & -\frac{\omega^4}{4k^2}
		\left[(g_0+h_0)^2c_1 - (g_0-h_0)^2 c_2 \right]
	\end{bmatrix}
	 	\langle {X}\rangle,
	 	\label{eq:M2direct2}	 	
\end{equation}
\end{widetext}
with the $c_i$s defined as above. The validity condition of the expansion is the same as in the previous sub-section.

As in Ref.~\cite{Dujardin2021}, the MI of stochastic origin is related to the growth rate of the second moment. The eigenvalues of the matrix in Eq.~\eqref{eq:M2direct2} can be written analytically. Their form is rather complicated: in general we have two complex conjugate eigenvalues ($\lambda_\pm $) with negative real part and one positive real eigenvalue, $\lambda_0$.
The MI gain is thus defined as
$
G_2(\omega) \equiv \frac{\lambda_0}{2}
$.
Since $G_2(\omega)$ is small for small $N_0$, we proceed like in Ref.~\cite{Abdullaev1996} to derive the following approximation 
\begin{equation}
G_2(\omega) \approx \frac{4 (\gamma P)^2 k \omega^4\left[8 c_2 k^3+ c_3 \omega^4 \left(c_1 j_0^2 - 4 c_2(\gamma P)^2 \right)\right]}{64 k^6 - 32 c_3 (\gamma P)^2 k^3 \omega^4 - \omega^8\left(c_1 j_0^2 - 4 c_2(\gamma P)^2 \right)},
\label{eq:G2_M2LP}
\end{equation}
with $j_0 \equiv g_0 + h_0$. Nevertheless, this expression is still very cumbersome and, below, we rely only on the numerically computed eigenvalue $\lambda_0$.

We note that the cumulant expansion could be extended to fourth order (the process being Gaussian, the third-order terms vanish), but the resulting terms are very involved and do not clarify the behavior of gain at large $\omega$, where the method breaks down (see below).

\subsection{Functional approach}

An alternative approach follows \cite{Gitterman2005,GittermanNoisy} and generalizes the use of Furustu-Novikov-Shapiro-Loginov formulas, on which the treatment of white noise in Ref.~\cite{Abdullaev1996} is based.

Let us consider one of the second moments $X_i$, $i=1,2,3$. They are functionals of $\delta\!g$. Since $\delta\!g\propto\chi$, its autocorrelation function has the form of Eq.~\eqref{eq:Rchi}. According to Ref.~\cite{Shapiro1978}, we have
\begin{equation}
 \langle\delta\!g \frac{\ud X_i}{\ud z}\rangle = \left(\frac{\ud}{\ud z} + B\right)\langle\delta\!g X_i\rangle.
 \label{eq:FNSLdiff}
\end{equation}

Two steps are needed to write an averaged system: (i) average directly Eq.~\eqref{eq:MIeqs2}; (ii) multiply each row  by $\delta\! g$ and average. We introduce three new variables, $X_{3+i} \equiv \delta\!g X_i$ and, in order to truncate the system, we assume $\langle\delta\! g^2 X_i\rangle = \sigma_{\delta\!g}^2 \langle X_i\rangle$, where $\sigma_{\delta\!g}^2 \equiv N_0 B \frac{\omega^4}{16}$ is the variance of the process. This last assumption is rigorously valid only for a dichotomous process \cite{GittermanNoisy} and in general an infinite hierarchy of equations is obtained for a Gaussian one, see \cite{Shapiro1978}. 

We obtain a 6\textsuperscript{th}-order system of ODEs
\begin{equation}
  \frac{\ud}{\ud z}
	 	\langle {X}\rangle= 
	 	\begin{bmatrix}
		0 & 0  & -2g_0  & 0 & 0 & -2\\
		0& 0  &2 h_0 & 0 & 0 & 2 \\
		h_0 & -g_0 & 0 & 1& -1& 0\\
		0 & 0 & -2\sigma_{\delta\!g}^2 & -B & 0 & -2 g_0 \\
		0 & 0 & 2\sigma_{\delta\!g}^2 &  0 & -B & 2 h_0 \\
		\sigma_{\delta\!g}^2 & -\sigma_{\delta\!g}^2 & 0 & h_0 & -g_0 & -B \\
	\end{bmatrix}
	 	\langle {X}\rangle.
	 	\label{eq:M2gitterman}	 	
\end{equation}
The matrix in Eq.~\eqref{eq:M2gitterman} has six eigenvalues. We observse numerically that one is real and negative, one ($\lambda_0$) is real and positive, the last four are two pairs of complex conjugate values with negative real part. The MI gain is defined as above, $
G_2(\omega) \equiv \frac{\lambda_0}{2}
$. 
By numerical inspection, we notice that the system in Eq.~\eqref{eq:M2gitterman} generally gives  different eigenvalues with respect to Eq.~\eqref{eq:M2direct2}. We stress that, in contrast to Sec.~\ref{ssec:cumexp2}, the accuracy of the functional approach does not require any condition on $\Eps$.
However, in the  white noise limit of $B\to\infty$, the system reduces to three independent variables \cite{GittermanNoisy} and we obtain the Eqs.~(26-30) of \cite{Abdullaev1996}, which as expected coincide with the cumulant expansion results.

In the next subsection, we will show what are the limits of validity of the two approximations and which fits better to numerical results.

\subsection{Results}
\label{ssec:LPnum}

\begin{figure}[hbtp]
\centering
\includegraphics[width= 0.4\textwidth]{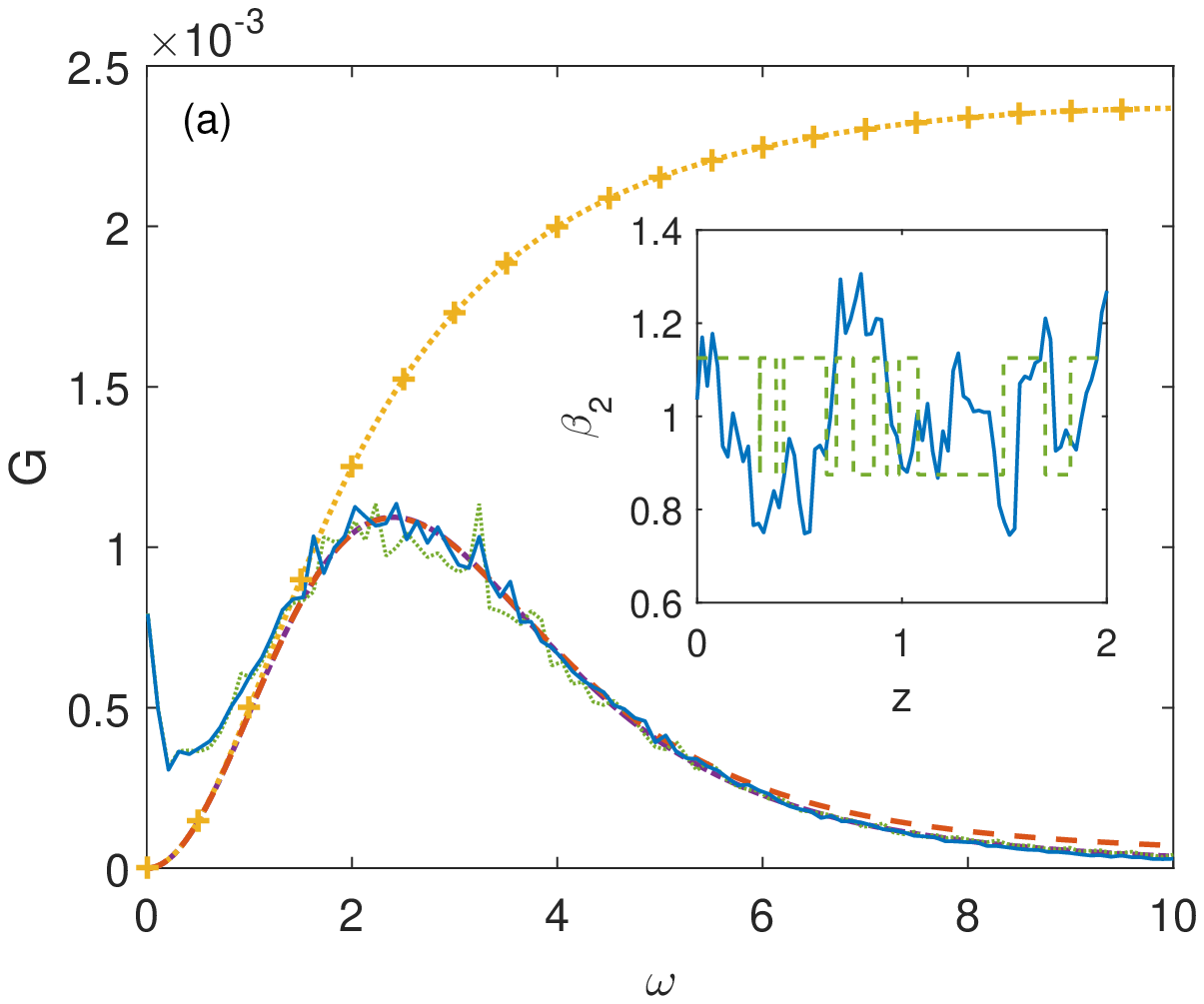}
\includegraphics[width= 0.4\textwidth]{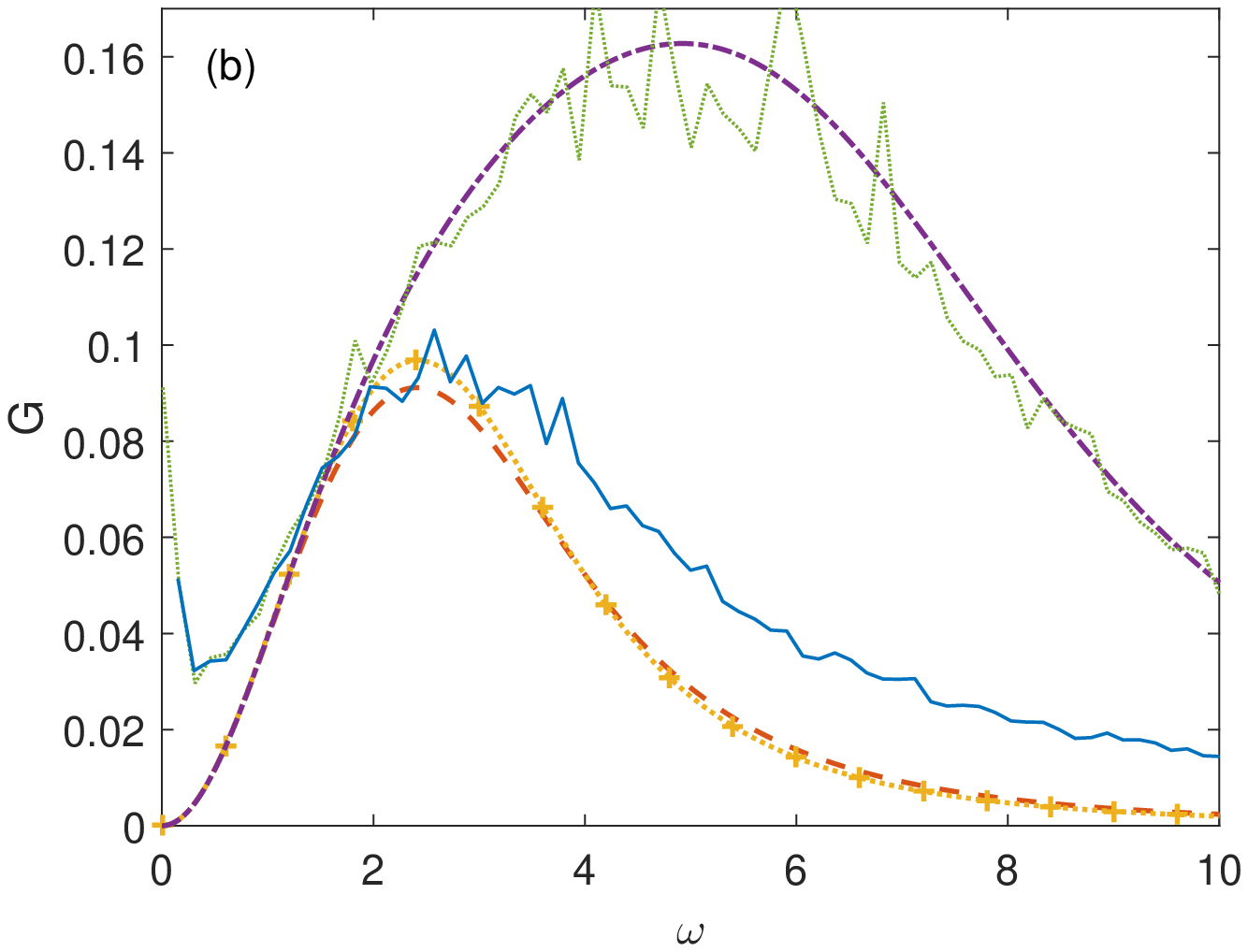}
\caption{MI gain as a function of detuning $\omega$ for a LP random dispersion.
Comparison of numerical values  for a Gaussian (blue solid lines) and dichotomous process  (green dotted lines) obtained from  Eq.~\eqref{eq:MIeqs1} with the estimates provided by  Eq.~\eqref{eq:M2direct2} (red dashed lines), by Eq.~\eqref{eq:M2gitterman}	(dash-dotted purple lines). We include also the gain corresponding to the white process (yellow dotted line with pluses). In panel (a) $N_0= 0.005$ and $B=4\pi$, so that $\varepsilon<0.5$  over the considered $\omega$ range, (b) $N_0= 0.4$ and $B=4\pi$, so that $\varepsilon\approx 1$ around the maximum gain.   The inset in panel (a) shows a single realization of the two processes.}
\label{fig:LPgain}
\end{figure}

In order to validate our theoretical analysis, we resort to solving Eq.~\eqref{eq:MIeqs1} by generating a large number, $N$, of realizations of $\chi$ and studying the moments of the resulting sample of solutions. The solution of Eq.~\eqref{eq:MIeqs1} is known exactly  for piecewise constant $\beta_2(z)$ in terms of transfer matrices \cite{Agrawal2012}.  For the Gaussian process, we choose the length $L$ of the domain sufficiently large to ensure that the sampling rate in the spatial frequency, \textit{i.e.}, $\Delta\!\kappa = \frac{2\pi}{L}$, fairly represents the PSD of the process, namely $B/\Delta\!\kappa \ll 1$. An array of identically distributed Gaussian random variables of zero mean and variance $\sigma^2 = \frac{N_0}{ 2\Delta\! z}$, with $\Delta\! z$ the sampling distance in $z$ (over which the dispersion is assumed constant), is generated numerically. This array is transformed to the $\kappa$ domain by means of FFT, then  multiplied by the Lorentzian filter of Eq.~\eqref{eq:LorLP} and transformed back to space domain. 
A realization of the dichotomous process is instead obtained in the $z$ domain, by randomly generating the switching distances (where $\chi$ changes sign) from a randomly generated exponential distribution of mean $2/B$.  Between two switching points the GVD is assumed constant.

We solve Eq.~\eqref{eq:MIeqs1} for a given initial condition $(x_1(0),x_2(0))^\mathrm{T} = (1,0)$, corresponding to balanced sidebands, over each random sequence of constant dispersion segments, to obtain a set of output vectors $(x_1(L),x_2(L))^\mathrm{T}$ from which we estimate the MI gain. Let $P_\mathrm{out} \equiv x_1^2(L)+x_2^2(L)$.
We then compute the mean gain, defined as \cite{Dujardin2021,Farahmand2004}
\begin{equation}
 G \equiv \frac{1}{2L}\log \langle \frac{P_\mathrm{out}}{P_\mathrm{in}}\rangle.
\end{equation} 

In all our results we take $\gamma P = \beta_2^0 = 1$. This is equivalent to putting Eq.~\eqref{eq:NLS1} in the standard adimensional form. 

In Fig.~\ref{fig:LPgain} we consider two examples of MI gain curves. We notice that the sidelobes exhibit a single maximum $G_\mathrm{max}$ at detuning $\omega_\mathrm{max}$. 

We consider two different values of $N_0$. For small values, the gain is generally small, see Fig.~\ref{fig:LPgain}(a), where $N_0=0.005$. We choose a long domain $L=5000$ to prevent finite-size effects that may appear at small $\omega$, and $\Delta\!z = 0.01$ for the Gaussian case. Satisfactory statistics are obtained for $N=2000$.

In this case $\varepsilon\approx0.5$ at $\omega=10$ (see the definition of $\varepsilon$ in \eqref{eq:defvarepsilon}), the cumulant expansion is thus expected to be valid. The functional approach gives very similar results, apart from some deviations in the large $\omega$ tails. The numerical data show that the two processes provide the same trend, that matches almost perfectly with the analytical estimates (the functional approach proves more accurate in the tails, as expected). The behavior near $\omega=0$ is due to the above-mentioned finite-size effects. We notice that the LP process provide a much narrower and smaller gain than the white noise (dotted yellow line with pluses). In this regime, the correlations of the stochastic process suppress the MI gain.

Then we study the case of an intermediate value $N_0=0.4$, where $\varepsilon>1$.
We choose $L = 50$, $\Delta\!z =0.01$, and $N =1000$. We show in Fig.~\ref{fig:LPgain}(b) how the numerical results compare to the analytical estimates. The functional approach proves very accurate for the dichotomous process over the full range of $\omega$. The cumulant expansion models the numerical data quite well, except in the large $\omega$ tails. We notice that the white noise gain is smaller than the LP gain lobes, particularly for the dichotomous case. Thus, for large $N_0$, correlations can improve the MI gain. 

\begin{figure}[hbtp]
\centering
\includegraphics[width=0.4\textwidth]{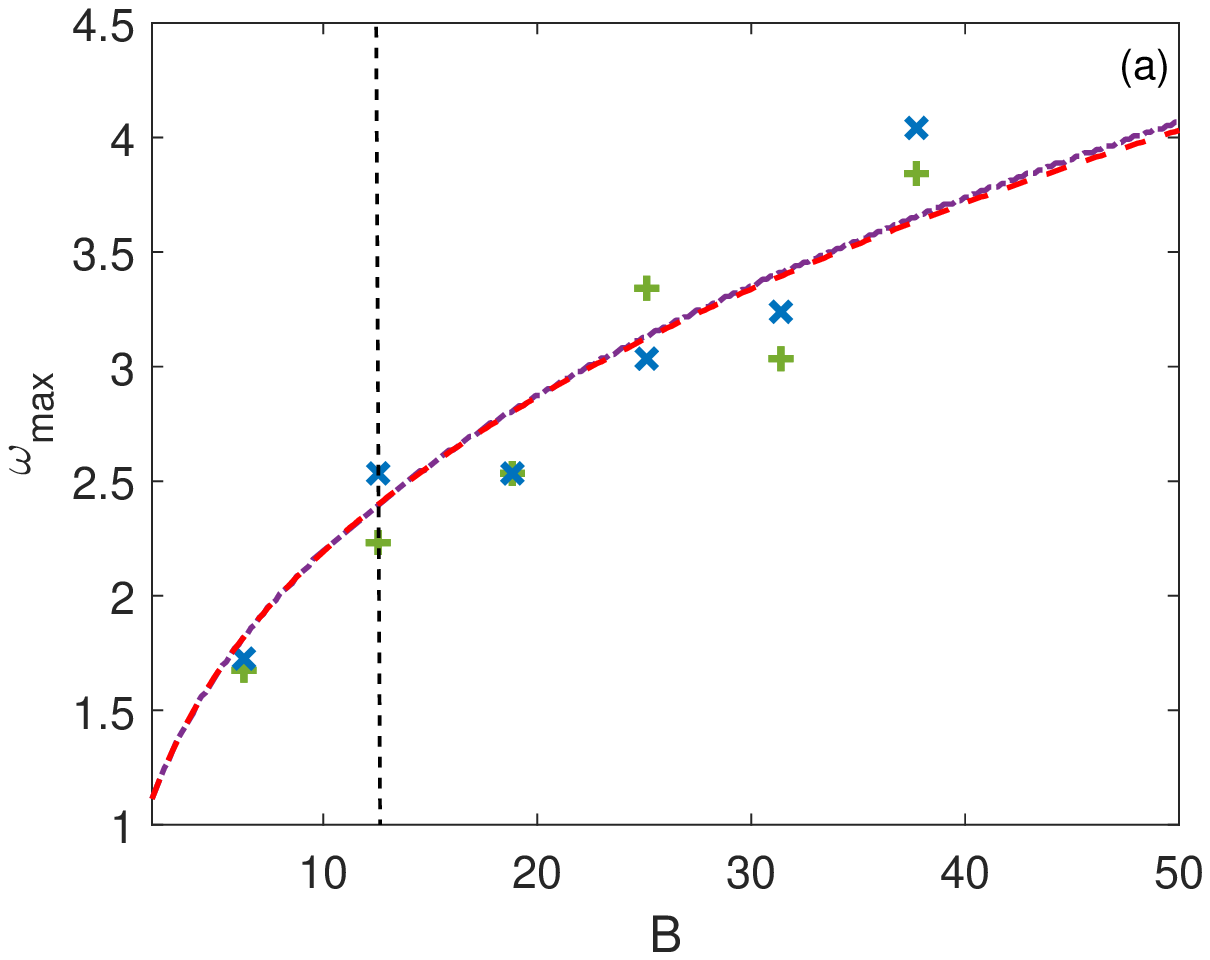}
\includegraphics[width=0.4\textwidth]{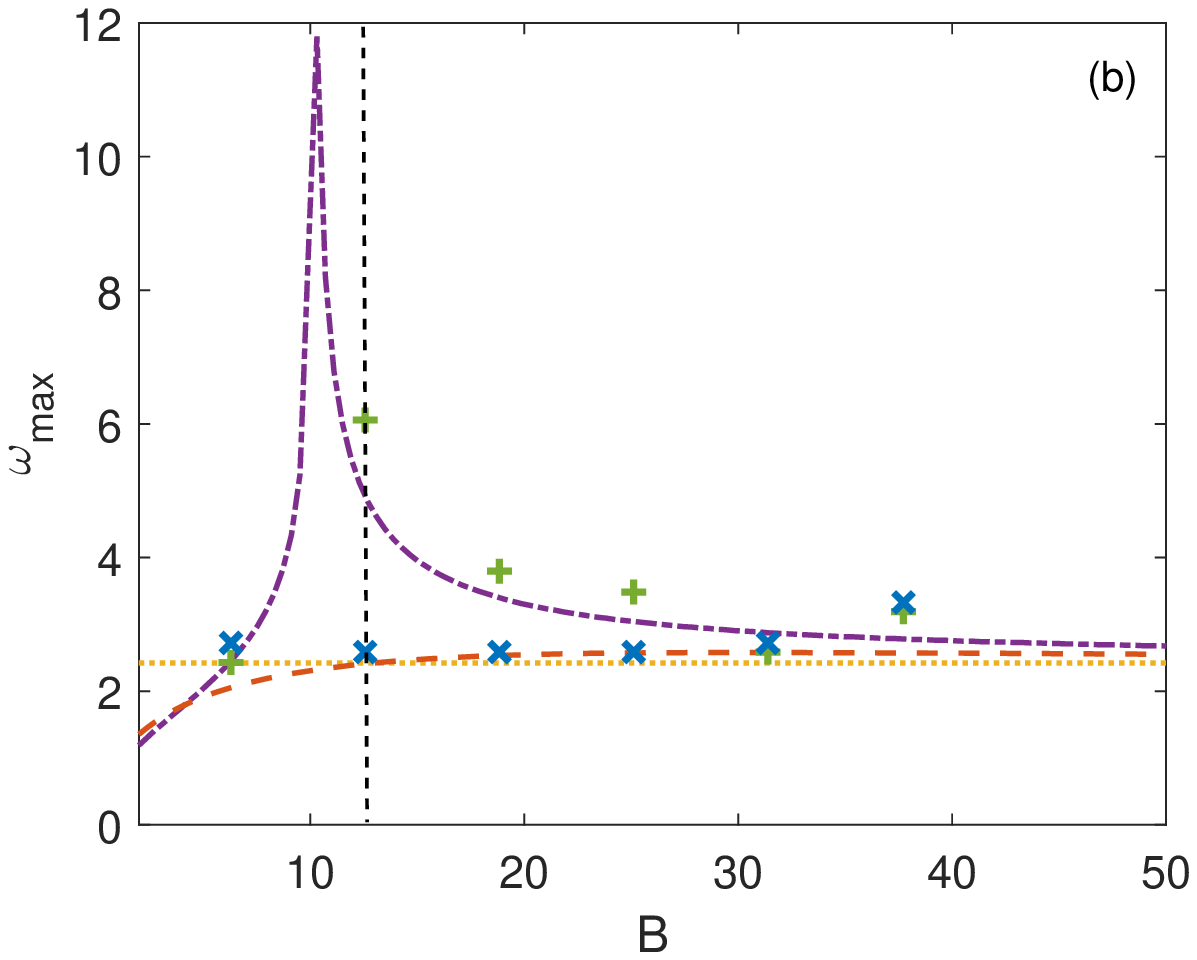}
\caption{Detuning at the gain maximum as a function of $B$. Numerical results of Gaussian  (blue crosses) and dichotomous (green pluses) process are compared to Eq.~\eqref{eq:M2direct2} (red dashed line), Eq.~\eqref{eq:M2gitterman} (purple dash-dotted line). We include $\omega_\mathrm{max}$ for the white noise, too, as reference (yellow dotted line). The dashed vertical lines highlight the value of $B$ used in Fig.~\ref{fig:LPgain}. (a) $N_0=0.005$, (b)  $N_0=0.4$. }
\label{fig:omegamaxLP}
\end{figure}
\begin{figure}[hbtp]
\centering
\includegraphics[width=0.4\textwidth]{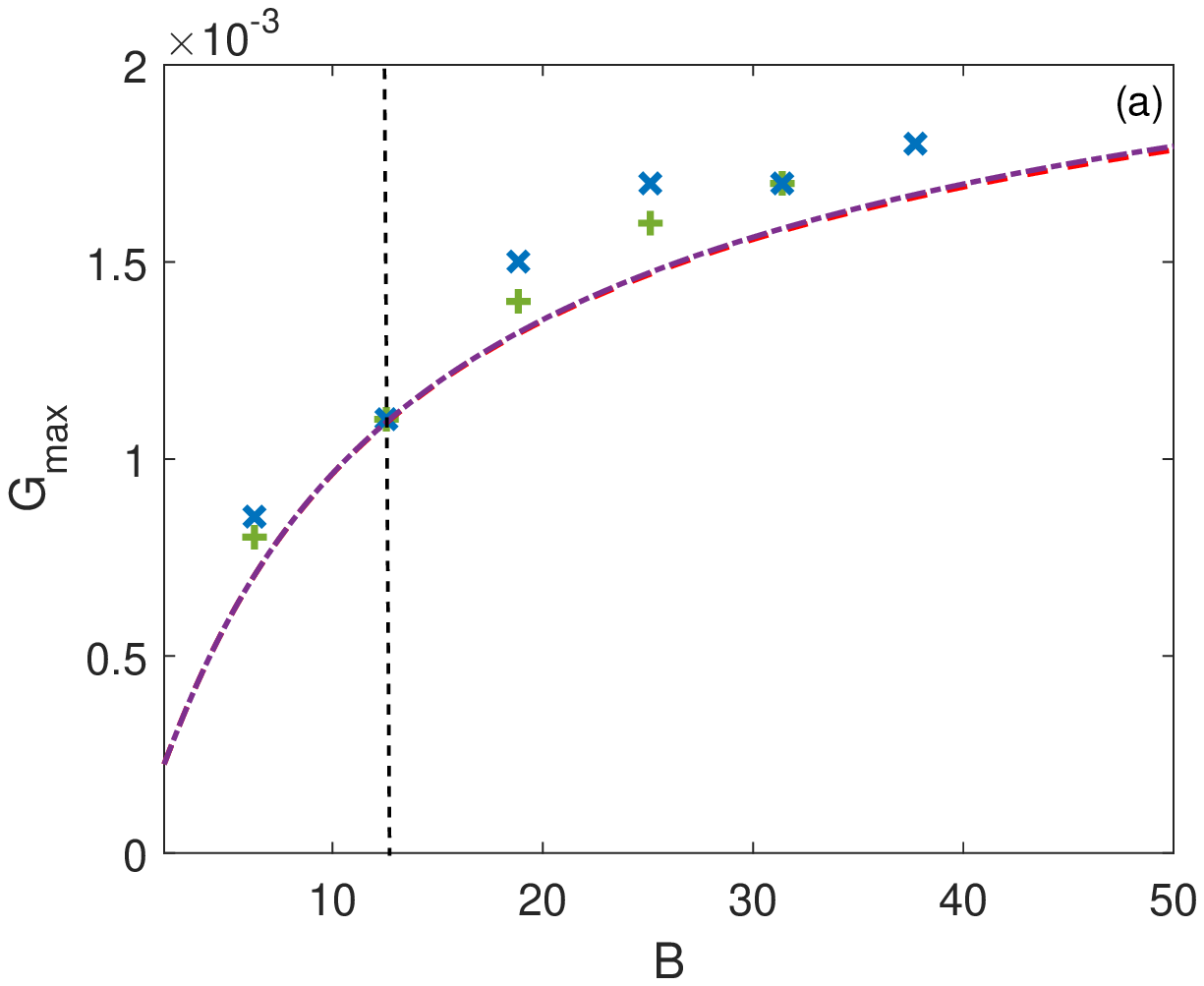}
\includegraphics[width=0.4\textwidth]{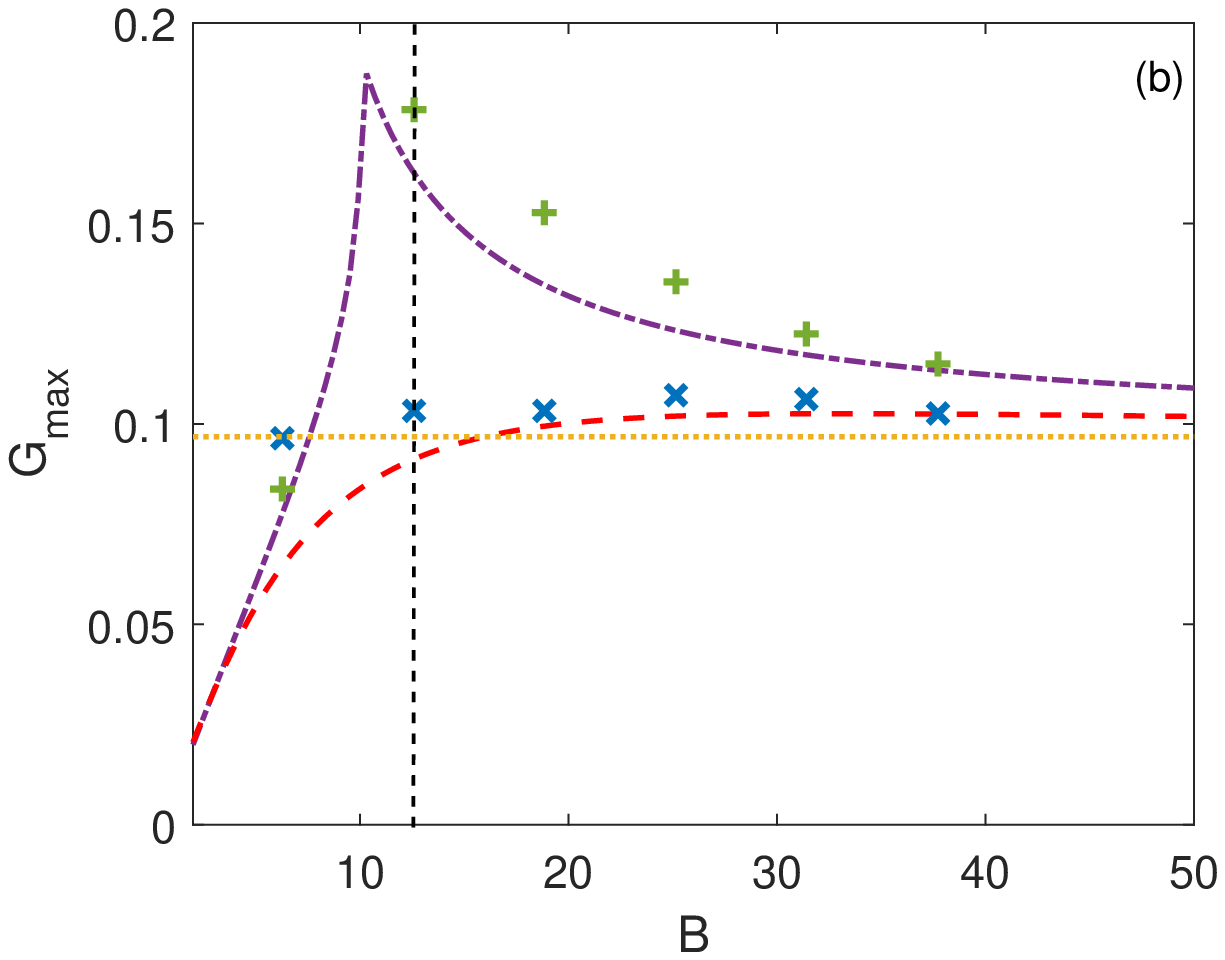}
\caption{Maximum gain as a function of $B$. Same convention as in Fig.~\ref{fig:omegamaxLP}.}
\label{fig:GmaxLP}
\end{figure}

In Figs.~\ref{fig:omegamaxLP}-\ref{fig:GmaxLP} we show the effect of variations of $B$ on  $\omega_\mathrm{max}$ and $G_\mathrm{max}$, respectively; $L$, $\Delta\!z$, and $N$ are chosen to guarantee a satisfactory statistical sample for each point.
 

For $N_0 = 0.005$  [Figs.~\ref{fig:omegamaxLP}(a) and \ref{fig:GmaxLP}(a)] the position as well as  the value of the maximum gain are, for both stochastic processes, in the same range of values and are, within the residual oscillation margins, well approximated by both methods. Both $\omega_\mathrm{max}$ and $G_\mathrm{max}$
 are monotone increasing functions of $B$. As $B\to+\infty$, they converge from below to the corresponding values for a white noise process. The limiting values are not shown, because they are well beyond the axis scales. 

The behavior for $N_0=0.4$ is very different [Figs.~\ref{fig:omegamaxLP}(b) and \ref{fig:GmaxLP}(b)]. First, for the dichotomous case, the numerical points are well approximated by the eigenvalues of Eq.~\eqref{eq:M2gitterman}. The maximum MI gain exhibits a sharp peak, see Fig.~\ref{fig:GmaxLP}(b). 
We notice in Fig.~\ref{fig:omegamaxLP}(b) that $\omega_\mathrm{max}$ diverges for this value of $B$: this means that the gain lobe never decays.

To understand this phenomenon, we study the eigenvalues of Eq.~\eqref{eq:M2gitterman} at $\omega\to\infty$ and we notice that, for $\sigma_{\delta\!g}=g_0$, i.e., $B=4\left(\beta_2^0\right)^2/N_0$, $G_2(\omega)$ converges, for small $N_0$, to $G_\mathrm{max} \approx \frac{1}{4}\left[-B+\sqrt{\frac{B^2+B\sqrt{B^2+64(\gamma P)^2}}{2}}\right]$.
Notice also that for this value $\sigma^2_\chi=1$, i.e., in dimensional units the GVD switches from $0$ to $2\beta_2^0$. In the present case, $B=10$ and $G_\mathrm{max}\approx 0.17$, as observed in Fig.~\ref{fig:GmaxLP}(b). For $N_0>1$, the reasoning is still valid: the value of $B$ for which  $G_2(\omega)$ does not decay at $\omega\to\infty$ is well predicted, while its value is generally larger than the approximated $G_\mathrm{max}$ above. From a physical point of view, this case is very pathological because higher-order dispersion effects should be included in Eq.~\eqref{eq:NLS1}.

For the Gaussian process, $\omega_\mathrm{max}$ and $G_\mathrm{max}$ stay close to the white noise limit (yellow dotted line) even for small $B$ and are satisfactorily described by the cumulant expansion for large $B$. As expected, the cumulant expansion is not accurate for small $B$ and is very different from the estimate of the functional approach.

We conclude that the cumulant expansion provides an accurate approximation for $G_\mathrm{max}$ and $\omega_\mathrm{max}$ for the Gaussian process. We recall that, on the contrary, this is not the case for the behavior of the MI gain at large $\omega$, see Fig.~\ref{fig:LPgain}(b). 


In conclusion, the LP processes yield a very small gain for $\varepsilon\ll 1$, always less than  white noise limit. For $\varepsilon\sim 1$, $G_\mathrm{max}$ can yield an MI gain larger than the white noise gain, but is about one order of magnitude smaller than the conventional MI gain in homogeneous fibers with anomalous GVD. 


\section{Narrowband random dispersion}
\label{sec:BP}

Here we consider the modulated process $\xi$ with the autocorrelation function in Eq.~\eqref{eq:Rxi}, for which a new degree of freedom, $\kappa_0$, is included. We start directly from Eq.~\eqref{eq:MIeqs2}.

\subsection{Direct cumulant expansion}
\label{ssec:BPDCE}
The cumulant expansion gives the same result as Eq.~\eqref{eq:M2direct2} with 
\begin{equation}
\begin{aligned}
c_1&=\frac{N_0}{2}\frac{B^2}{B^2+\kappa_0^2} ,
\\
c_2 &=  \frac{N_0B^2}{4} \left[\frac{1}{B^2 + (2k-\kappa_0)^2} + \frac{1}{B^2 + (2k+\kappa_0)^2}\right],\\
c_3 &=N_0 B\kappa \frac{B^2-\kappa_0^2+4k^2}{\left[B^2 + (2k-\kappa_0)^2\right]\left[B^2 + (2k+\kappa_0)^2\right]}. 
\end{aligned}
\label{eq:ciBP}
\end{equation}
We observe that, while $c_1$ is constant,  $c_2$ and $c_3$ exhibit a resonant lineshape behavior close to $2k=\kappa_0$, \textit{i.e.}, the 1\textsuperscript{st} PR condition: the former exhibits a maximum at $4k^2\approx \kappa_0^2-\frac{B^4}{4 \kappa_0^2}$, while the latter crosses zero at $4k^2= \kappa_0^2-B^2$.
We recall that the $m$-th order PR condition is in general $2k = m \kappa_0$, which gives 
\begin{equation}
\omega_{\mathrm{PR},m}^2 = \sqrt{4\left(\frac{\gamma P}{\beta_2^0}\right)^2+\left(\frac{m \kappa_0}{\beta_2^0}\right)^2} - \frac{2\gamma P}{\beta_2^0} .
\label{eq:omegaPR}
\end{equation} 

For $B\ll\kappa_0$, we can set $c_1\approx 0$. Then around the first parametric resonance detuning, tedious but straightforward calculations show that the instability gain is approximated by
\begin{equation}
G_2(\omega) \approx \frac{1}{2}\frac{(2\gamma P)^2\omega^4}{ \kappa_0^2}c_2,
\label{eq:G2_M2} 
\end{equation}
that attains its maximum approximately at $\omega_\mathrm{PR,1}$, 
\begin{equation}
\begin{aligned}
G_2^\mathrm{max}\approx G_2(\omega_\mathrm{PR,1})
&\approx \frac{N_0(\gamma P)^2}{2}\frac{\omega_\mathrm{PR,1}^4}{\kappa_0^2} \\
&= \frac{N_0(\gamma P)^2}{\left(2\beta_2^0\right)^2}
\frac{\omega_\mathrm{PR,1}^2}{\omega_\mathrm{PR,1}^2+\frac{4\gamma P}{\beta_2^0}}.
\end{aligned}
\label{eq:G2_M2cummax}
\end{equation}

We remark also that Eq.~\eqref{eq:G2_M2} is composed of a Lorentzian factor depending on the process PSD ($c_2$) and a factor independent of the process. Moreover, it turns out that the maximum MI gain, Eq.~\eqref{eq:G2_M2cummax}, coincides, for small $\omega$ and a given $N_0$ with the white noise MI gain evaluated at $\omega=\omega_{{\rm PR},1}$, as hinted at in Ref.~\cite{Zhang1992}.


Notice that in  Eq.~\eqref{eq:G2_M2cummax}, the MI gain is proportional to $\frac{N_0\omega_\mathrm{PR,1}^4}{\kappa_0^2}$, while the PR-MI scales like $\theta\frac{\omega_\mathrm{PR,1}^2}{\kappa_0}$ \cite{Armaroli2012}, with $\theta$ the (constant) amplitude of the periodic variation. As in the low-pass case, according to Eq.~\eqref{eq:G2_M2cummax} the maximal MI gain depends mainly on $N_0$. For $B\to 0$, the analogy to PR would suggest, instead, a dependence on the amplitude of the fluctuations in real space, i.e., $N_0 B$, as discussed also in Ref.~\cite{Zhang1992}. Below, we clarify this ostensible inconsistency.

\subsection{Near-resonance reduction}

In analogy to Ref.~\cite{Zhang1992}, a significant simplification of Eq.~\eqref{eq:MIeqs1} can be obtained by the conventional averaging method used for PR \cite{Landau1976}.

In order to average Eq.~\eqref{eq:MIeqs1}, we let 
\begin{equation}
\begin{aligned}
x_1(z) &= y_1(z)\cos\left(\frac{\kappa_0}{2} z\right) + y_2(z)\sin\left(\frac{\kappa_0}{2} z\right) \\
x_2(z) &= \frac{\kappa_0}{2g_0}\left[y_1(z)\sin\left(\frac{\kappa_0}{2} z\right) - y_2(z)\cos\left(\frac{\kappa_0}{2} z\right)\right],
\end{aligned}
\label{eq:avetrans}
\end{equation}
where $x_2$ is written assuming $g=g_0$ constant, in the spirit of the variation of constants. 

By averaging out oscillating terms, noticing that $\frac{\delta\! g}{g_0} = \xi$, and employing the phase-quadrature representation of $\xi$, see Eq.~\eqref{eq:PQxi}, we obtain
\begin{equation}
\begin{aligned}
   	\kappa_0 \dot y_1 & = \Delta^2 y_2 +  \Gamma\psi_1 y_2 - \Gamma\psi_2 y_1
   	\\
    \kappa_0 \dot y_2 & = -\Delta^2 y_1 +  \Gamma\psi_1 y_1 + \Gamma\psi_2 y_2,
\end{aligned}
\label{eq:MIeqs_res}
\end{equation}
with $\Gamma\equiv \frac{1}{2}\left(\frac{\kappa_0^2}{4}-g_0^2\right)$ and $\Delta^2 \equiv k^2-\frac{\kappa_0^2}{4}$ quantifies the detuning from the PR condition; close to resonance, $\Delta^2 = (k+\kappa_0/2)(k- \kappa_0/2) \approx \kappa_0 (k-\kappa_0/2)$.
 
Notice that the PR-MI gain of the first PR tongue is obtained from Eq.~\eqref{eq:MIeqs_res} by putting $\psi_1=\psi_2=\theta$. It reads
\begin{equation}
G_\mathrm{PR,1} = \frac{\sqrt{2(\Gamma \theta)^2 - \Delta^2}}{\kappa_0}.
\label{eq:GPR1}
\end{equation}

Starting from Eq.~\eqref{eq:MIeqs_res}, in this section we only study the evolution of second moments.
We introduce $Y_1 \equiv y_1^2$, $Y_2 \equiv y_2^2$, and $Y_3 \equiv y_1y_2$. It is easy to verify that $ {Y}\equiv (Y_1,Y_2,Y_3)^\mathrm{T}$ obeys 
\begin{equation}
	 	 	\kappa_0\frac{\ud}{\ud z}
	 	 {Y} = 
	 	\begin{bmatrix}
	 		-2\Gamma\psi_2 & 0& 2\Delta^2+2\Gamma\psi_1 \\
	 		0 & 2\Gamma\psi_2 & -2\Delta^2  + 2\Gamma\psi_1 \\
	 		-\Delta^2+\Gamma\psi_1 & \Delta^2 +\Gamma\psi_1 & 0
	 	\end{bmatrix}
	 	 {Y},
	 	\label{eq:MIeqs2_res}
\end{equation}
where the parameters are the same as those used throughout this section. 

Two random processes appear in Eq.~\eqref{eq:MIeqs2_res} and both the cumulant expansion and functional approach need generalizing accordingly. While we present the former in App.~\ref{app:cumNR}, because it provides very similar results to the direct cumulant expansion of the previous subsection, the latter is reported below.

\subsection{Functional approach near resonance}

Following \cite{Burov2016}, we generalize the functional approach. 
We  define $Y_{3+i} \equiv \psi_1 Y_i$, $Y_{6+i} \equiv \psi_2 Y_i$, and $Y_{9+i} \equiv \psi_1 \psi_2 Y_i$, $i=1,2,3$. We perform four different averaging steps: (i) average directly Eq.~\eqref{eq:MIeqs2_res}, (ii) multiply each row of Eq.~\eqref{eq:MIeqs2_res} by $\psi_1$ and average, (iii) multiply  by $\psi_2$ and average; (iv) multiply by $\psi_1\psi_2$ and average.
We employ the formula of differentiation in Eq.~\eqref{eq:FNSLdiff} and its generalization
\begin{equation}
\begin{gathered}
 \langle\psi_1\psi_2 \frac{\ud Y_i}{\ud t}\rangle = 
 \left(\frac{\ud}{\ud t} + 2B\right)\langle\psi_1\psi_2  Y_i\rangle.
 \end{gathered}
 \
\end{equation}

If we assume as above that we can factor the variance out if the same process occurs twice in an average, we obtain

\begin{equation}
  \kappa_0\frac{\ud \langle {Y}\rangle}{\ud z}
	 	= 
	 	\left[
	 	\begin{array}{c|c|c|c}
	 	     A_2 & \Gamma C_2' & \Gamma C_2'' & \mathbf{0}\\
	 	    \hline
	 	    \sigma^2_\xi \Gamma C_2' &  A_2-B \mathbf{I}  & \mathbf{0} & \Gamma C_2'' \\
	 	    \hline
	 	    \sigma^2_\xi  \Gamma C_2'' & \mathbf{0} &   A_2-B \mathbf{I} & \Gamma C_2' \\
	 	    \hline
	 	 \mathbf{0} & \sigma^2_\xi \Gamma C_2'' & \sigma^2 _\xi\Gamma C_2'&  A_2-2B \mathbf{I}
	 	\end{array}
	 	\right]
	 	\langle {Y}\rangle,
	 	\label{eq:M2PRgitterman}	 
\end{equation}
with 
\begin{equation}
\begin{gathered}
{A_2} = \begin{bmatrix}
	 		0 & 0& 2\Delta^2 \\0& 0& -2\Delta^2\\
	 		-\Delta^2 & \Delta^2  & 0
	 	\end{bmatrix}, \\
	{C_2'} = \begin{bmatrix}
	 		0 & 0 & 2 \\0 & 0 & 2\\
	 		1 & 1 & 0
	 	\end{bmatrix},  \;
 {C_2''} =  \begin{bmatrix}
	 		-2 & 0 & 0  \\0 & 2 & 0 \\ 0 & 0 & 0
	 	\end{bmatrix},  	 	
\end{gathered}
\label{eq:A2C2BP}
\end{equation}
and denoting $\mathbf{0}$ and $\mathbf{I}$  the null and identiy matrix, respectively. 

We compute numerically the eigenvalues of the matrix in Eq.~\eqref{eq:M2PRgitterman} and look for the dominant one, $\lambda^*$.
The maximum MI gain, for $\Delta^2=0$, is  found analytically as
\begin{equation}
(G_2^\mathrm{PR})^\mathrm{max}  = \frac{1}{4\kappa_0} \left[\sqrt{B^2+4 N_0 B \omega_\mathrm{PR,1}^4}-B\right].
 \label{eq:G2_M2Gittmax}
\end{equation}

We remark that the dependence on $\sigma_\xi$ is different from Eq.~\eqref{eq:G2_M2cummax}. For small $B$, the gain is no longer proportional to $\frac{\omega_\mathrm{PR,1}^4 N_0 }{\kappa_0^2}$, but to   $\frac{\omega_\mathrm{PR,1}^2 \sqrt{N_0 B} }{\kappa_0}$, as we would formally obtain in the conventional periodic dispersion case, once we replace the amplitude of parametric oscillation $\theta$ with  $\sqrt{2}\sigma_\xi$.

Thus a very different result is found when compared to the cumulant expansions above. We now assess which approximation works better by comparing them to numerical solutions of Eq.~\eqref{eq:MIeqs1}.

\subsection{Results}
\label{ssec:BPnum}

\begin{figure}[hbtp]
\centering
\includegraphics[width= 0.4\textwidth]{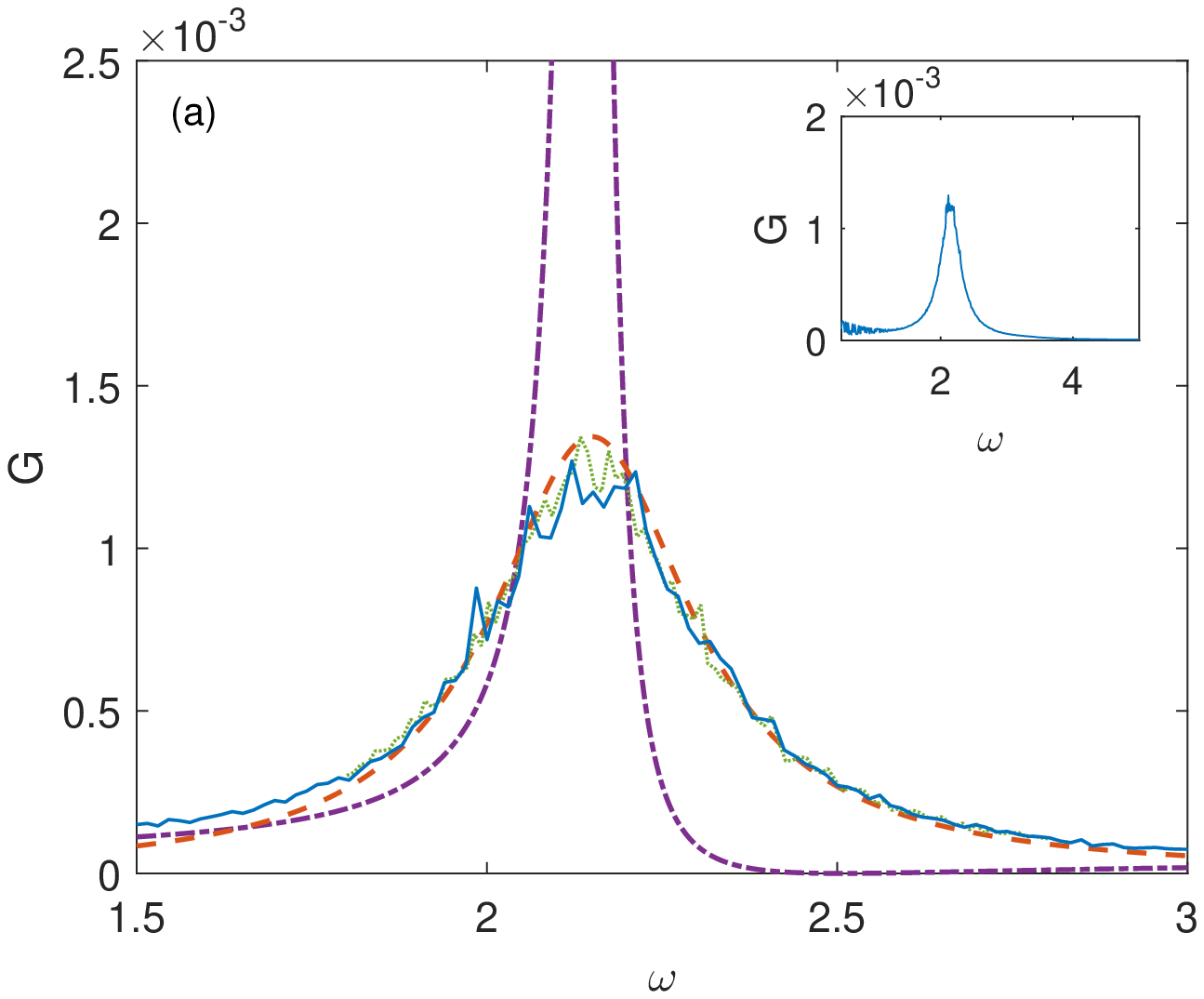}
\includegraphics[width= 0.4\textwidth]{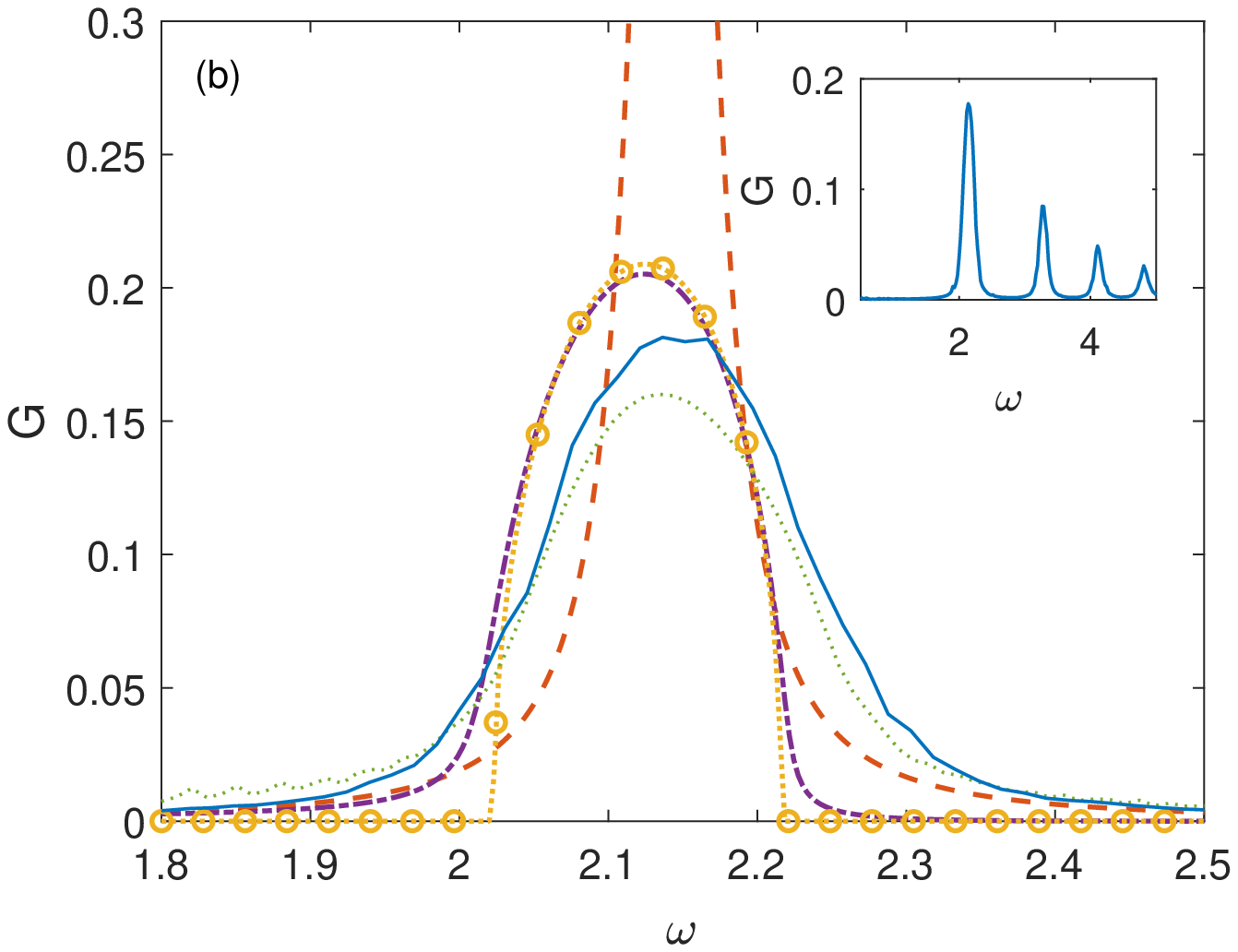}
\caption{MI gain as a function of detuning $\omega$ for a BP random dispersion, $\kappa_0=2\pi$.
Comparison of numerical values for a Gaussian (blue solid lines) and dichotomous process  (green dotted lines) obtained from  Eq.~\eqref{eq:MIeqs1} with the estimates provided by  Eqs.~\eqref{eq:M2direct2} and \eqref{eq:ciBP}  (red dashed lines), and by Eq.~\eqref{eq:M2PRgitterman}	(dash-dotted purple lines). In panel (a) $N_0= 0.005$ and $B=\pi/4$, so that $\varepsilon<0.1$ around the PR resonance, (b) $N_0= 3.2$ and $B=\pi/32$, so that $\varepsilon\gg 1$ around the $\omega_\mathrm{max}$. In panel (b), we include the PR-MI gain of Eq.~\eqref{eq:GPR1} with $\theta=\sqrt{2}\sigma_\xi=\sqrt{\pi/5}$ (yellow dotted line with circles). The insets show the numerical results for a Gaussian process on a larger $\omega$ range. }
\label{fig:BPgain}
\end{figure}

In order  to generate a realization of the process $\xi$ with Gaussian distribution (resp.~dichotomous), we employ the same approach in the spectral (resp.~spatial) domain as above, generate two independent LP processes and modulate them according to Eq.~\eqref{eq:PQxi}. The numerical domain is obviously discretized in both cases: for the Gaussian process a lower limit on $L$ is required as in Sec.~\ref{ssec:LPnum}, the dichotomous process requires a short $\Delta\!z$ to avoid spurious correlations. We do not forget to correctly sample the period $\Lambda_0$ of the process spatial oscillations. 

While in the previous section we were interested in the limit $B\to\infty$ to contrast the LP processes to white noise, here we aim at understanding the opposite limit, $B \to 0$,  with fixed and finite $N_0 B$.
We expect that the gain associated to the stochastic fluctuations converges to the one of the periodically modulated fiber in this limit.

First, we consider a small perturbation, $N_0 = 0.005$, with a intermediate bandwidth $B=\pi/4$ around $\kappa_0=2\pi$. We take $L=500$, $\Delta\! z=0.01$; a consistent statistical sample is collected for $N=2000$. In Fig.~\ref{fig:BPgain}(a) we compare the numerical mean gain $G(\omega)$ with the analytical estimates obtained above. For such a small perturbation only a single MI peak can be observed (see the inset for a larger detuning range). We notice that the MI gain is centered about $\omega_ \mathrm{PR,1}$ with gain $G_\mathrm{max}$ and width $\Delta\!\omega$. The Gausssian and dichotomous processes give two almost identical results. The cumulant expansion provides a very good approximation, because $\varepsilon\approx 0.1$. The near-resonance functional approach proves instead very imprecise, as far as both $G_\mathrm{max}$ and $\Delta\!\omega$ are concerned. We explain this as follows: it is easy to verify that the period of the modulation is comparable to correlation length, $\Lambda_0\approx \zeta_c$. Thus, we apply successively two distinct averaging procedures (near-resonant expansion and functional approach) upon two perturbations occurring at the same scale as if they were independent. This is a sure recipe for failure.  

Moreover, we are not showing here the result of Eq.~\eqref{eq:G2R_M2} that coincide with the direct cumulant expansion around $\omega_\mathrm{PR,1}$, but is skewed towards $\omega\to 0$, contrary to the numerical results. 

As a second case, we consider a large $N_0=3.2$ and a small bandwidth $B=\pi/32$. All the other  parameters  are the same as the previous. In Fig.~\ref{fig:BPgain}(b) we observe, as above, that the MI lobe occurs around $\omega_ \mathrm{PR,1}$ and the Gaussian and dichotomous processes provide two very close results. For such a large perturbation, several MI sidelobes can be observed, in analogy to PR-MI, see the inset of Fig.~\ref{fig:BPgain}(b). Now $\varepsilon\approx 6.3 \gg 1$; we thus expect that the cumulant expansion fails to correctly describe the numerical results: indeed it overestimates the peak MI while it underestimates its width (red dashed line). It captures approximately the behavior of the MI gain in the tails (both left-hand and right-hand sides). The functional method provides instead a good approximation (purple dash-dotted  line). For these parameters, Eq.~\eqref{eq:M2PRgitterman} provides a result very close to the conventional PR gain (yellow dotted line with circles) apart from the tails. We can thus state that for $\zeta_c\gg \Lambda_0$, the proposed functional approach gives a good estimate of the numerically estimated mean gain, because the two independent approximations are performed in the correct order on the two different scales, i.e., for $B\ll\kappa_0$, the BP process can be considered a small perturbation to the PR-MI effect.

\begin{figure}[hbtp]
\centering
\includegraphics[width=0.4\textwidth]{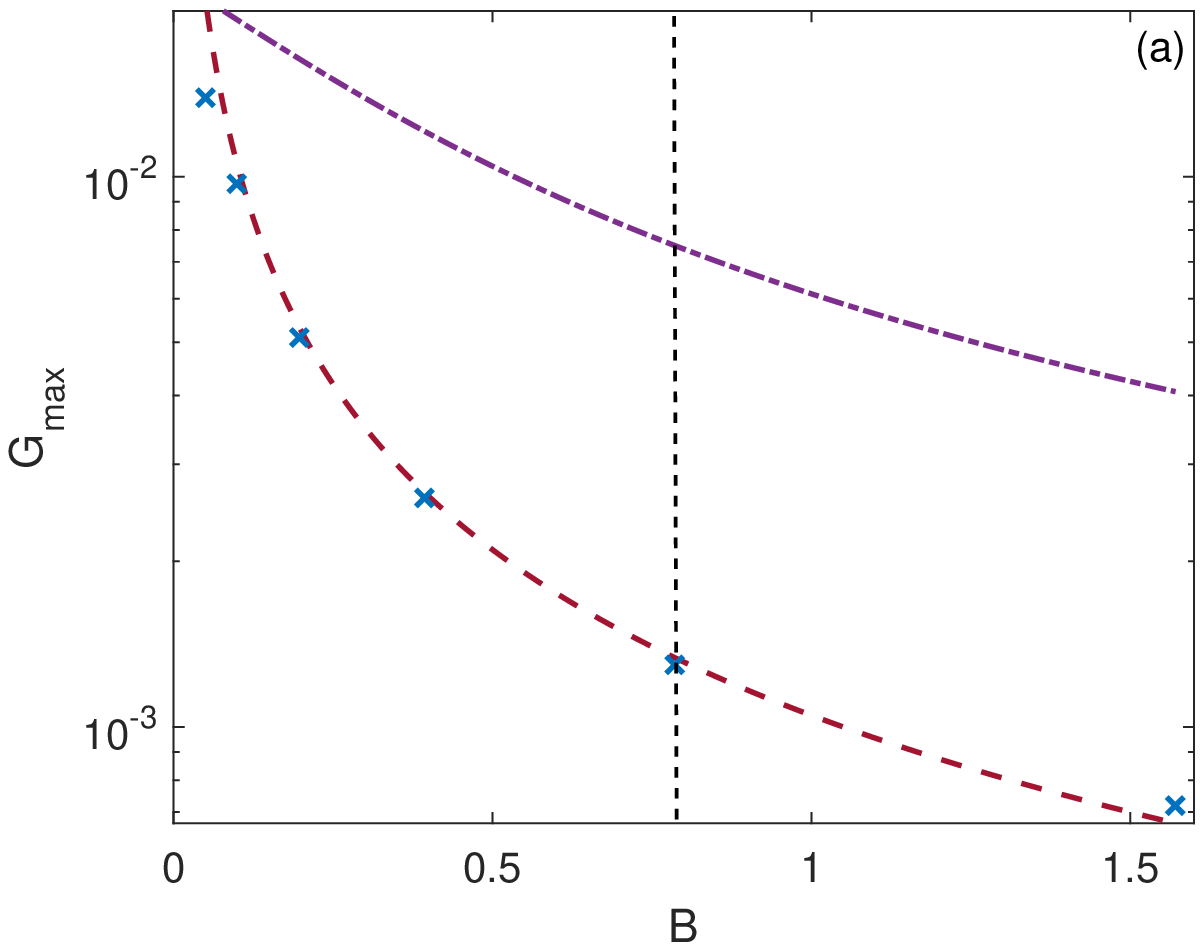}
\includegraphics[width=0.4\textwidth]{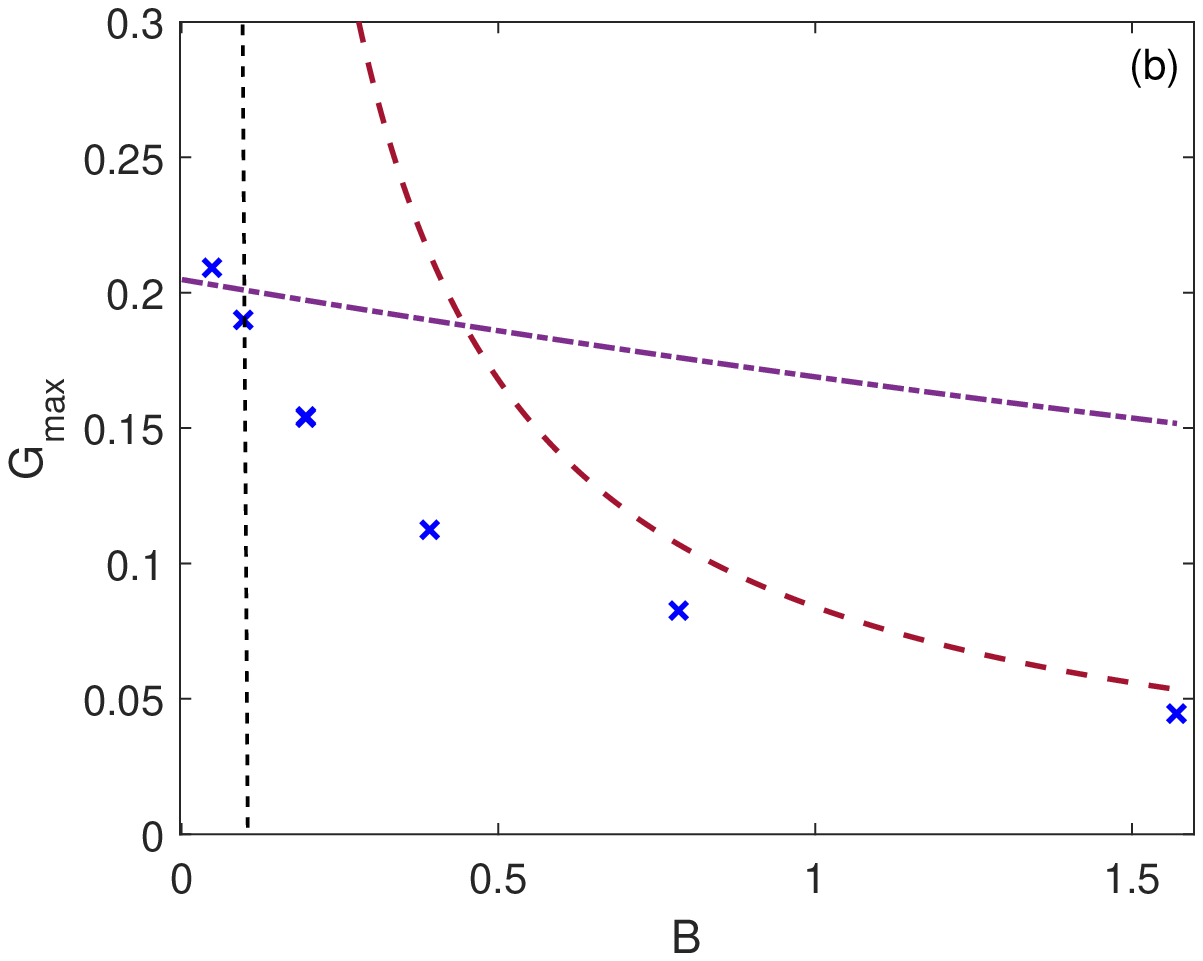}
\caption{Maximum gain $G_\mathrm{max}$ as a function of $B$ at constant $N_0B$. Same convention as in Fig.~\ref{fig:omegamaxLP}, dichotomous process excluded. We take a constant $\kappa_0=2\pi$ and (a) $N_0B = 0.0039$, (b) $N_0B=\pi/10= 0.314$. Notice the ordinate axis is in logarithmic scale in panel (a). The dashed vertical lines highlight the values of $B$ used in Fig.~\ref{fig:BPgain}(a)[(b)], respectively.}
\label{fig:GmaxBP}
\end{figure}
 We show in Fig.~\ref{fig:GmaxBP} the effect of variations of $B$ on $G_\mathrm{max}$. We keep $\kappa_0$ and $N_0 B$  constant; $L$, $\Delta\!z$, and $N$ are chosen to guarantee a significant statistical sample for each value. Only the Gaussian process is considered. 

For small $N_0B=0.0039$ the MI gain follows very well the cumulant expansion: $G_\mathrm{max}$ grows for $B\to 0$ (notice the logarithmic scale). The functional approach always overestimate it. 
For larger  $N_0B=\pi/10$ we still observe a decreasing trend of $G_\mathrm{max}$. It is apparent that the cumulant expansion is valid only for large $B>\pi/2$ and completely loses its validity below $B=\pi/4$. The functional approach works well in the PR limit, i.e., for $B<\pi/16$. Indeed, $\zeta_c$ depends inversely on $B$ and the approximation is expected to become more faithful. The residual discrepancies may depend on a systematic error of the averaging procedure or on numerical inaccuracies.  In brief, we show the transition from the stochastic regime, where the MI gain scales with $N_0$, to the parametric regime where 
it depends on $\sigma_\xi$, i.e. the fluctuation amplitude.

\begin{figure}[hbtp]
\centering
\includegraphics[width=0.4\textwidth]{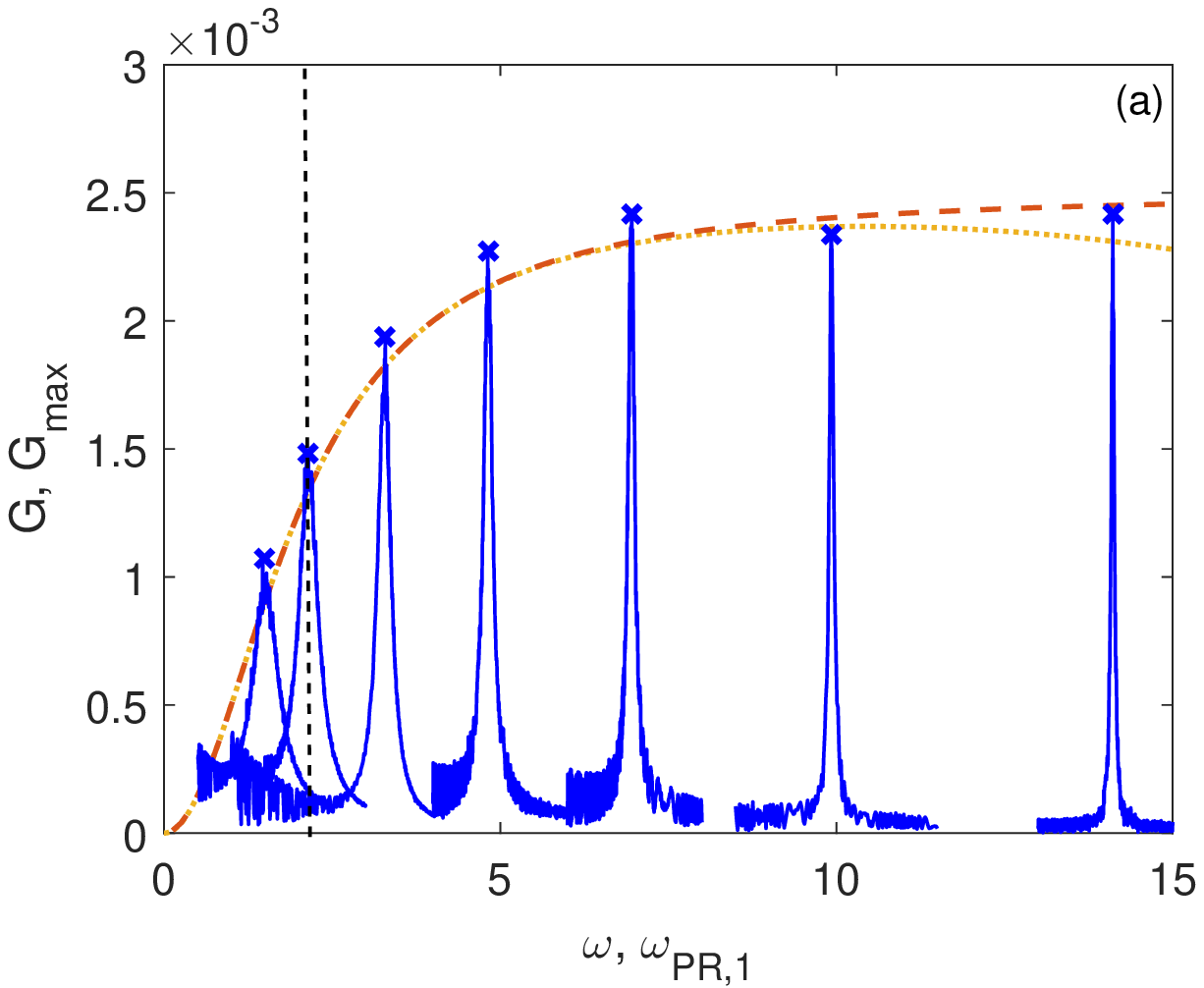}
\includegraphics[width=0.4\textwidth]{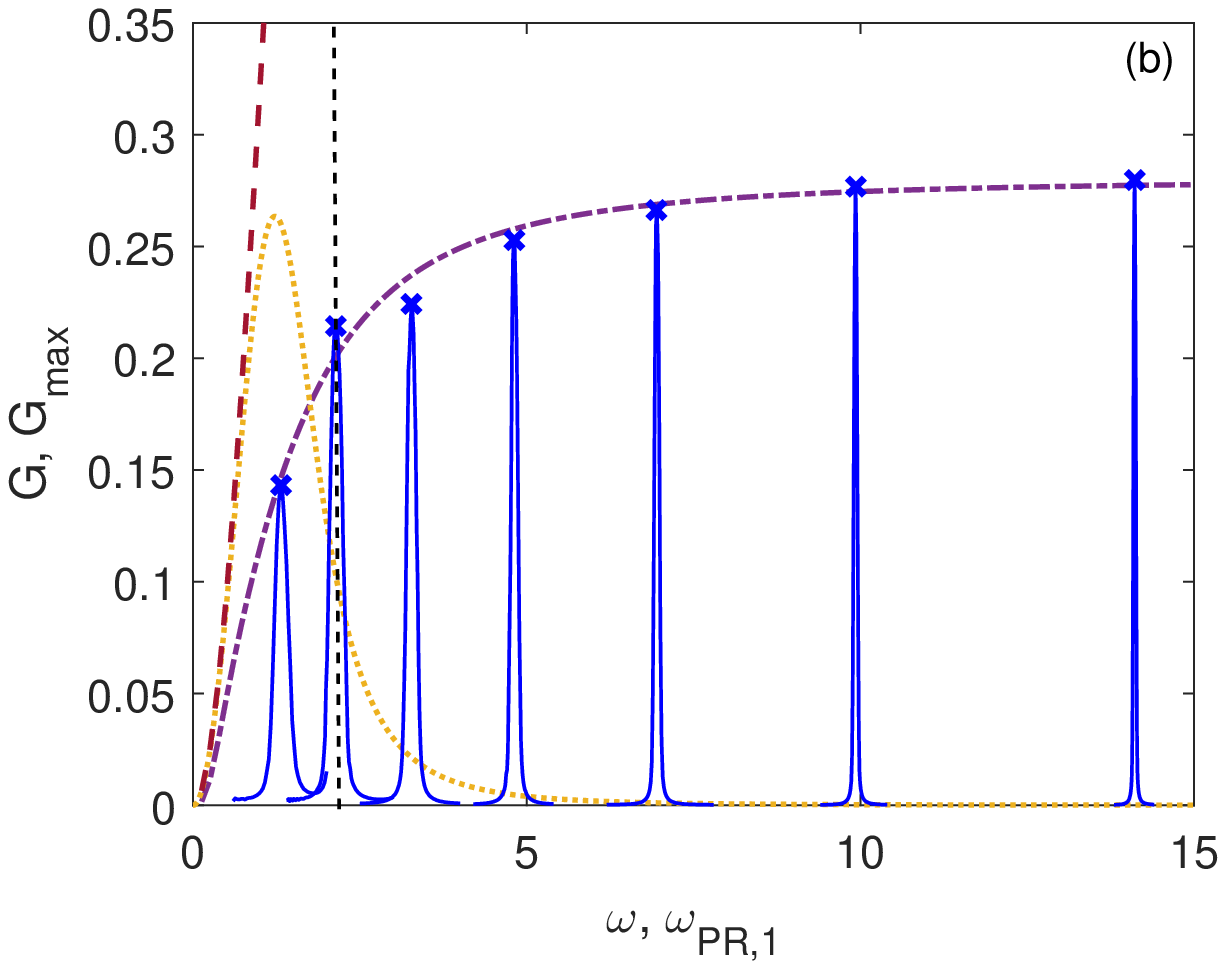}
\caption{MI gain curves $G(\omega)$ and their maxima  $G_\mathrm{max}$
obtained for different values of $\kappa_0\in\{\pi,2\pi,4\pi,8\pi,16\pi,32\pi,64\pi\}$. Comparisons of numerical results [blue crosses and blue solid lines] with the peak of Eq.~\eqref{eq:G2_M2cummax} (red dashed line), of Eq.~\eqref{eq:G2_M2Gittmax} [purple dash-dotted line, not shown in panel (a), out of scale]. We include also $G_2(\omega)$ for the white noise limit at the corresponding (yellow dotted line). The dashed vertical lines highlight the value of $\kappa_0$ used in Figs.~\ref{fig:BPgain} and \ref{fig:GmaxBP}. (a) $N_0 = 0.005$, $B=\pi/4$; (b) $N_0 = 3.2$, $B=\pi/32$.  }
\label{fig:GmaxBPvskappa0}
\end{figure}

Finally, we report in Fig.~\ref{fig:GmaxBPvskappa0} different MI gain curves obtained from numerical simultations of the Gaussian process by varying $\kappa_0\in\{\pi,2\pi,4\pi,8\pi,16\pi,32\pi,64\pi\}$.
The simulation parameters are chosen for each value to ensure that the sample is statically significant. 

In Fig.~\ref{fig:GmaxBPvskappa0}(a), we choose $N_0=0.005$ and $B= \pi/4$ and show that the MI lobes have a Lorentzian shape dominated by a simple envelope, as predicted by Eq.~\eqref{eq:G2_M2}. We explicitly show $G_2^\mathrm{max}$ obtained from Eq.~\eqref{eq:G2_M2cummax} for the values of $\omega_\mathrm{PR,1}$ corresponding to a continuous set of $\kappa_0$ (red dashed line) and compare it to the values of $G_\mathrm{max}$ (blue crosses). They match very well over the full range of $\omega$. The result of the functional approach, Eq.~\eqref{eq:G2_M2Gittmax}, is not shown here, because it always overestimates the MI gain. We also include the white noise gain for the given $N_0$ (yellow dotted line). For such a small PSD value, it exhibits a monotone growth up to $\omega\approx 9$, compare to Fig.~\ref{fig:LPgain}(a). The expression in Eq.~\eqref{eq:G2_M2cummax} coincides with it up to this point and tends to $N_0/2$ for $\kappa_0,\omega_\mathrm{PR,1}\to\infty$. This strengthens the idea that the MI gain of stochastic origin is ruled predominantly by $N_0$, at least for small values. 

If, instead, we choose $N_0=3.2$ and $B= \pi/32$, we observe, see Fig.~\ref{fig:GmaxBPvskappa0}(b), that the MI lobes are sharper than a Lorentzian. Their maxima (blue crosses) are well approximated by the functional approximation (purple dash-dotted line), while the cumulant expansion (red dashed line) always overestimates the MI gain by more than 3 times.
For such a small $B$, Eq.~\eqref{eq:G2_M2Gittmax} effectively coincides with Eq.~\eqref{eq:GPR1} (not shown) as expected. 
As  in the previous case, the gain monotonously increases with $\kappa_0$, consistently with the PR nature of the phenomenon, see Eq.~\eqref{eq:GPR1}. 
If we consider the white noise, we see that the MI gain exhibits a narrow lobe, that attains a maximum for $\omega\approx 1.25$ and coincides with \eqref{eq:G2_M2cummax} only for small detuning, as expected. 

As $\kappa_0$ is increased, the modulation period of the process becomes smaller, while $\zeta_c$ is kept constant; thus, the near-resonant approximation performs better for large $\kappa_0$ and the discrepancies observed  \ref{fig:GmaxBPvskappa0}(b) are probably due to a systematic error or residual numerical inaccuracies.

\section{Conclusions}

We discuss the modulational instability in nonlinear optical fibers in which the group-velocity dispersion is randomly modulated. In contrast to the exactly solvable case of white noise or of random kicks, we consider the case of stochastic processes with exponentially decaying autocorrelation function. This is equivalent to a Lorentzian-shaped power spectral-density, i.e., the process is \textit{colored}. Two families of colored processes are studied: low-pass and band-pass. 
For each family, we consider Gaussian and dichotomous processes.  

The distinction between the two stochastic processes turns out to be important in the LP  and marginal in the BP case. 

While for very small perturbation the LP yields very small MI gain, for larger power spectral densities it can yield MI sidelobes larger than the white noise, sitting in the same range of detuning and exhibiting larger gain in the tails. For small bandwidth, the gain disappears, while in the opposite limit it converges to the white noise limit. The variance demanded for obtaining a measurable MI gain is large, though, and the dichotomous process looks more promising in view on an experimental characterization of the phenomenon, because values of bandwidth (correlation length) exist where the gain is quite larger than the white noise limit. 

As far as a BP process is concerned, if the perturbation is large enough, we may observe several MI sidelobes sitting around PR detunings. We focus on the dominant first peak: it converges to the PR sidelobe for small bandwidth and is generally broader and smaller for small correlation lengths.

We compare our numerical results to different analytical approximations, based on the cumulant expansion (as formalized by van Kampen) or the functional (Furutsu-Novikov-Loginov-Shapiro) formulas. 

While the former is reliable only for small perturbations and small detuning and provides some qualitative estimates elsewhere, the latter provides good results  for the dichotomous processes, for which the closure of the system is rigorously obtained.

For both families of correlation functions, the functional method emerges as more reliable and allows us to describe the transition from parametric to stochastic resonances in the BP case. Notwithstanding, the cumulant expansion provides good estimates of the tails of resonant peaks even for relatively small bandwidth values (large correlation length), beyond the expected validity range of the approximation. 


Our results pave the way for tailoring MI gain sidebands in optical fibers by means of stochastic GVD fluctuations and suggest the  regimes to achieve that.
Such fluctuations can be implemented by continuous or discrete variations of fiber specifications. 


\vskip 1cm
\section*{Acknowledgments}
The work was supported in part the by the French government through the Programme Investissement d’Avenir with the Labex CEMPI (Grant ANR-11-LABX-0007-01) and the I-SITE ULNE (Grant ANR-16-IDEX-0004 ULNE, projects VERIFICO)
managed by the Agence Nationale de la Recherche. The work was also supported by the Nord-Pas-de-Calais Regional Council and the European Regional Development Fund through the Contrat de Projets \'Etat-R\'egion (CPER) and IRCICA.


\appendix

\section{Cumulant expansion near resonance}
\label{app:cumNR}
In this section we will apply the cumulant expansion to Eq.~\eqref{eq:MIeqs2_res}. 

The cumulant expansion to second order is built by decomposing Eq.~\eqref{eq:MIeqs2_res} as $\kappa_0\dot Y = \left[A_2 + \Gamma \psi_1 C_2' +\Gamma \psi_2 C_2''\right] Y$, with $A_2$, $C_2$, and $C_2'$ as in Eq.~\eqref{eq:A2C2BP}.

As $\psi_{1,2}$ are mutually independent, the expansion is obtained by computing $K_2$ separately for $C_2'$ and $C_2''$, according to Eq.~\eqref{eq:CumExp2}, and adding them up.

We obtain
\begin{equation}
  \kappa_0\frac{\ud \langle {Y}\rangle }{\ud z}
	 	= 
	 	\begin{bmatrix}
		6 d_2 \Gamma^2 & 2 d_2 \beta^2 &   2\Delta^2-4\Gamma^2 d_3 \\
        2 d_2 \beta^2 & 6 d_2 \beta^2 &   -2\Delta^2 +4\Gamma^2 d_3  \\		
        2 \Gamma^2 d_3  - \Delta^2 & -2 \Gamma^2 d_3 + \Delta^2 & 4 \Gamma^2 d_2
	\end{bmatrix}
	 	\langle {Y}\rangle,
\label{eq:M2res2}
\end{equation}
with 
\begin{equation}
\begin{aligned}
d_2&\equiv   \frac{1}{\kappa_0} \int \limits_0^{\infty}\ud \zeta R_{\psi_1}(\zeta) \cos \frac{2\Delta^2}{\kappa_0} \zeta = \frac{N_0 B^2}{2\kappa_0} \frac{1}{B^2 + \frac{4\Delta^4}{\kappa_0^2}}\\
d_3 &\equiv \frac{1}{\kappa_0}\int \limits_0^{\infty}\ud \zeta R_{\psi_1}(\zeta) \sin \frac{2\Delta^2}{\kappa_0} \zeta = \frac{N_0 B}{2 \kappa_0} \frac{\frac{2\Delta^2}{\kappa_0}}{B^2 + \frac{4\Delta^4}{\kappa_0^2}}.
\end{aligned}
\end{equation} 
We note that there is here no counterpart to $c_1$ and near PR $\kappa_0 d_2\approx 2c_2$.

The dominant eigenvalue of the matrix in Eq.~\eqref{eq:M2res2} is \textit{exactly} $\lambda^* = 8 d_2 \Gamma^2$, so the instability gain is
\begin{equation}
G_2^\mathrm{PR} =  \frac{8d_2\Gamma^2}{\kappa_0},
\label{eq:G2R_M2}
\end{equation}
which attains a maximum  $(G_2^\mathrm{PR})^\mathrm{max} \approx \frac{N_0\omega_\mathrm{PR,1}^4}{2\kappa_0^2}$, identical to what found above in Eq.~\eqref{eq:G2_M2cummax}.

\end{document}